\documentclass[12pt]{iopart} 
\usepackage[pdftex]{graphicx} 
\usepackage{enumerate} 
\usepackage{braket} 

\usepackage{color} 
\usepackage{cite} 
\usepackage{iopams}  

	\definecolor{aaltoGray}{RGB}{146,139,129}%
	\definecolor{aaltoBlue}{RGB}{0,101,189}%
	\definecolor{aaltoRed}{RGB}{237,41,57}%
    \definecolor{aaltoGreen}{RGB}{0,155,58}%
	\definecolor{aaltoOrange}{RGB}{255,121,0}%

\begin{document}

\title[Multi-scale model for the structure of hybrid perovskites]{Multi-scale model for the structure of hybrid perovskites: Analysis of charge migration in disordered MAPbI$_3$ structures}

\author{Jari~J\"arvi$^{1,2}$, Jingrui~Li$^1$ and Patrick~Rinke$^1$}

\address{$^1$ Department of Applied Physics, Aalto University, P.O. Box 11100,
FI-00076 AALTO, Finland}
\address{$^2$ Department of Physics, University of Helsinki, P.O. Box 64,
FI-00014 University of Helsinki, Finland}
\ead{jari.jarvi@aalto.fi}
\vspace{10pt}
\begin{indented}
\item[]June 2018
\end{indented}

\begin{abstract}

We have developed a multi-scale model for organic-inorganic hybrid perovskites (HPs) that applies quantum mechanical (QM) calculations of small HP supercell models to large coarse-grained structures.  With a mixed quantum-classical hopping model, we have studied the effects of cation disorder on charge mobilities in HPs, which is a key feature to optimize their photovoltaic performance. Our multi-scale model parametrizes the interaction between neighboring methylammonium cations (MA$^+$) in the prototypical HP material, methylammonium lead triiodide (CH$_3$NH$_3$PbI$_3$, or MAPbI$_3$). For the charge mobility analysis with our hopping model, we solved the QM site-to-site hopping probabilities analytically and computed the nearest-neighbor electronic coupling energies from the band structure of MAPbI$_3$ with density-functional theory. We investigated the charge mobility in various MAPbI$_3$ supercell models of ordered and disordered MA$^+$ cations. Our results indicate a structure-dependent mobility, in the range of 50--66 cm$^2$V$^{-1}$s$^{-1}$, with the highest observed in the ordered tetragonal phase. 

\end{abstract}

%
\vspace{2pc}

\noindent \textit{This is the version of the article before peer review or
editing, as submitted by an author to the New Journal of Physics. IOP Publishing
Ltd is not responsible for any errors or omissions in this version of the
manuscript or any version derived from it. The Version of Record is available
online at} https://doi.org/10.1088/1367-2630/aae295.

%
%
\maketitle
%

\section{Introduction} \label{intro}

Hybrid organic-inorganic perovskites (HPs) are the primary candidates to substitute silicon as the cheap and efficient solar cell material of the future \cite{Green2014}.  Methylammonium lead triiodide CH$_3$NH$_3$PbI$_3$ (shortened to MAPbI$_3$ hereafter) is the spearhead of this novel photovoltaic (PV) materials class with an unprecedented development in its power conversion efficiency (PCE), from ca.~10\% in 2012 to the present 23\%, approaching the best silicon-based single-junction devices \cite{NRELchart}. However, the most efficient HPs are currently unstable due to structural degradation in atmospheric water \cite{Berhe2016}. Conversely, stable HPs are still too inefficient in their PCE.  New computational models, developed in parallel with experimental research, aim to improve the PCE of HPs via detailed understanding of the atomic structure and expedite the introduction of HPs into consumer-grade solar devices. While several mixed-halide HP compositions have recently entered the field, MAPbI$_3$ has maintained its status as the archetypal HP in the latest research.

MAPbI$_3$ combines the key properties of a good PV material --- a suitable band gap of ca.~1.6 eV \cite{Yamada2014}, good visible-light photon absorption \cite{DeWolf2014} and long diffusion length \cite{DongQ2015}. At present, the nature of the charge-carrier mobility in MAPbI$_3$ remains unclear. Recent studies suggest that the mobility is fairly modest in comparison to typical inorganic semiconductors, less than 100 cm$^2$V$^{-1}$s$^{-1}$ for electrons and holes \cite{Brenner2015,Karakus2015}. This indicates that the observed long diffusion length and the resulting high PV performance likely arise from a long charge-carrier lifetime. To optimize the charge mobility, we need detailed insight into the relationship between the mobility and the corresponding MAPbI$_3$ structure.

Recent research has focused on solving the atomistic origins of these properties, yet the structural complexity of MAPbI$_3$ and other HPs has hampered progress. Structural investigations have focused on the role of the organic methylammonium (MA$^+$) cations, specifically the rotational dynamics of the cations and their interplay with the inorganic PbI$_3$ lattice \cite{Mosconi2014,Frost2014b,Chen2015,Leguy2015,Lahnsteiner2016,Li2016,Li2017,Li2018a,Li2018b}. Depending on the cation orientation, coupling with the lattice via hydrogen bonding distorts the framework geometry and affects the electronic properties, for example by altering the nearest-neighbor (NN) atomic orbital overlap and the lattice vibrational modes. The exact relationship between this dynamical structure and the material's PV properties remains debated. Suggested mechanisms include for example charge-carrier pathways formed by ferroelectric domains of ordered MA$^+$ cations \cite{Frost2014a}, the effects of the lattice distortion on charge localization \cite{Kang2017} and potential fluctuation effects on the charge-carrier mobility due to dynamical MA$^+$ disorder \cite{Ma2017}. However, detailed computational studies of these processes require electronically accurate models, which significantly restricts the size of the studied systems.

To describe properties that derive from the atomic and electronic structure with sufficient accuracy, we need to incorporate quantum mechanical (QM) information. Methods such as density-functional theory (DFT) \cite{Hohenberg1964} deliver this QM insight, but are limited to relatively small structures ($<$100 unit cells).  However, due to the orientational freedom of the MA$^+$ cations in each unit cell, HPs will be disordered and require modeling on much larger lengths scales than accessible by DFT. We approach this accuracy-size conundrum with a new multi-scale model, which we have developed specifically for HP structures. Our model is based on a DFT parameterization of small supercell models and then applicable to large MAPbI$_3$ structures.

In our previous MAPbI$_3$ unit-cell energy-optimization study \cite{Li2016} we recognized two stable low-energy structures, corresponding to diagonal and face-to-face MA$^+$ orientations in the unit cell. We then parametrized the MAPbI$_3$ atomic structure by studying the MA$^+$s in neighboring unit cell pairs \cite{Li2017}. We focused on the MA$^+$ orientations, describing them by a C-N dipole vector, and defined the dipole pair configurations as ``pair modes'' (PMs hereafter, see section~\ref{pm_description}).  We analyzed the effect of energy optimization on the PM distribution in 33 $4 \times 4 \times 4$ MAPbI$_3$ supercells. In the optimized structures, we observed an absence of the diagonal dipoles and a notable change in the PM distribution, for example an increase in the number of perpendicular PMs. As a good approximation, the observed changes were also independent of the initial structures before optimization. These findings motivated our description of low-energy MAPbI$_3$ structures by a specific ``optimized'' PM distribution.  Moreover, it enabled us to generate new large-scale structures using the optimized PM distribution for further analyses, such as charge migration.

In this paper, we study the charge migration in large-scale MAPbI$_3$ structures with a semi-classical site-to-site hopping model. We generated the structures based on our previous observations in our supercell DFT studies, applying the PM description and a self-developed structure-generation algorithm.  We then analyzed the effect of dipole orientations on the charge migration velocity, by comparing the generated structures to various reference systems, for example ordered dipole configurations (corresponding to tetragonal and orthorhombic phases), parallel and antiparallel dipoles and randomly oriented dipoles.  We also analyzed the effect of structural defects (such as planar defects and precipitates) on the charge velocity. Finally, we estimated the charge mobility in our model based on an experimental reference value for the scattering time at room temperature \cite{Karakus2015}.

This paper is organized as follows. In section~\ref{model} we review the PM concept and introduce methods for the MAPbI$_3$ structure generation and the charge migration study. In section~\ref{results} we introduce and discuss the results, and then conclude our findings in section~\ref{conclusion}.

\section{Computational models and methods} \label{model}

To analyze the large-scale MA$^+$ cation order (or disorder), we generated new systems based on the PM distribution of the energy-optimized MAPbI$_3$ supercell models from our previous study \cite{Li2017}. We developed a semi-classical site-to-site hopping model to study charge migration in MAPbI$_3$ structures and compared the statistical migration velocities in various dipole configurations. In the following, we will first review the PM concept and then introduce the methods for the large-scale structure generation (see section~\ref{struct_gen}) and the charge migration study (see section~\ref{mig_model}).

\begin{figure}[ht] \centering
    \includegraphics[width=1.0\textwidth]{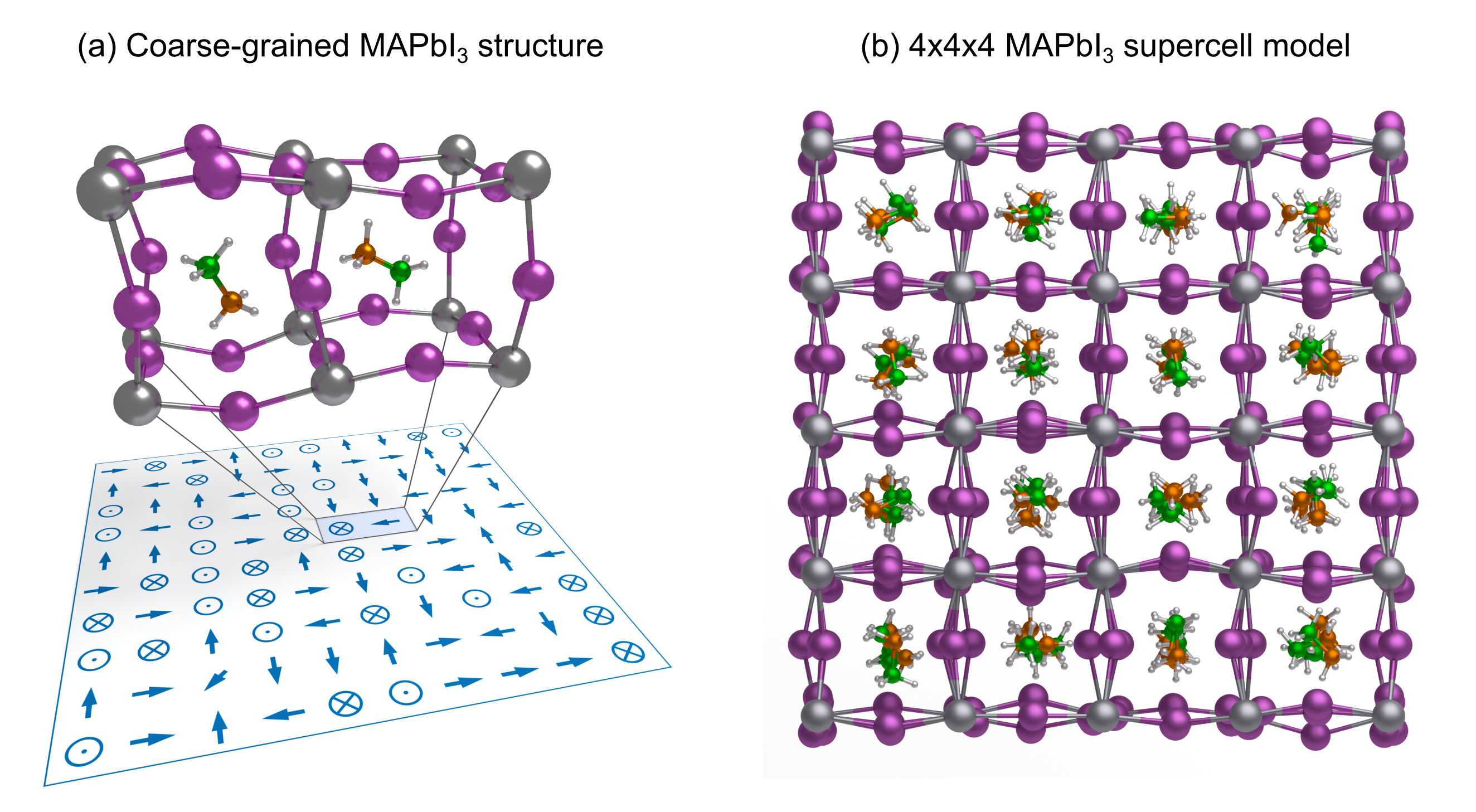} \caption{(a)
    Coarse-grained model of a MAPbI$_3$ structure, in which MA$^+$ cations are
    modelled as C-N dipole vectors, and the PbI$^-$ lattice deformation is
    parameterized into neighboring dipole pairs, the ``pair modes''.  (b) $4
    \times 4 \times 4$ supercell model of a DFT-optimized MAPbI$_3$ structure,
    showing the I displacement in the lattice. Pb, I, C, N and H atoms are
    colored in gray, purple, green, orange and white, respectively.}
\label{fig_coarse} \end{figure}

\subsection{Pair mode description} \label{pm_description}

We developed the PM concept based on our previous studies in MAPbI$_3$ structure optimization, using the all-electron numeric-atom-centered orbital DFT code FHI-aims \cite{Blum2009,Havu2009,Levchenko2015,Xinguo/implem_full_author_list}. In these studies we applied the Perdew-Burke-Ernzerhof (PBE) generalized gradient approximation \cite{Perdew1996} with Tkatchenko-Scheffler van der Waals (vdW) interactions \cite{Tkatchenko2009} as the exchange-correlation functional. Based on our single-unit-cell structure optimization study \cite{Li2016}, the MA$^+$ cation has 32 discrete stable orientations in a unit cell --- 8 diagonal and 24 face-to-face with a small deviation angle to the lattice plane. In these low-energy states the ammonium end of the cation forms hydrogen bonds with the lattice I$^-$ atoms, pulling them towards the cation.  The amount of displacement is the largest in the face-to-face MA$^+$ orientations, which subsequently has ca.~21 meV lower energy than the diagonal orientation. We have parametrized the structural energies of this framework distortion by describing the MA$^+$ cation as a dipole vector in the C-N bond direction and studying the dipole configurations in neighboring unit cell pairs, the PMs (see figure~\ref{fig_coarse}(a)).  The displacement of the I$^-$ atom depends not only on the dipole orientation in a single unit cell, but also on the orientation in the neighboring cell. By considering the rotational and reflectional symmetries of the dipole pairs, we were able to reduce the amount of $32^2 = 1024$ pairs to only 86 unique PMs. Furthermore, by approximating the face-to-face orientations in each unit cell to be identical (i.e. 4-fold degenerate without the lattice-plane deviation angle), the amount of unique PMs in this 14-orientation framework (with 8 diagonal and 6 face-to-face orientations) was reduced to 25.

\begin{figure}[ht] \centering
    \includegraphics[width=1.0\textwidth]{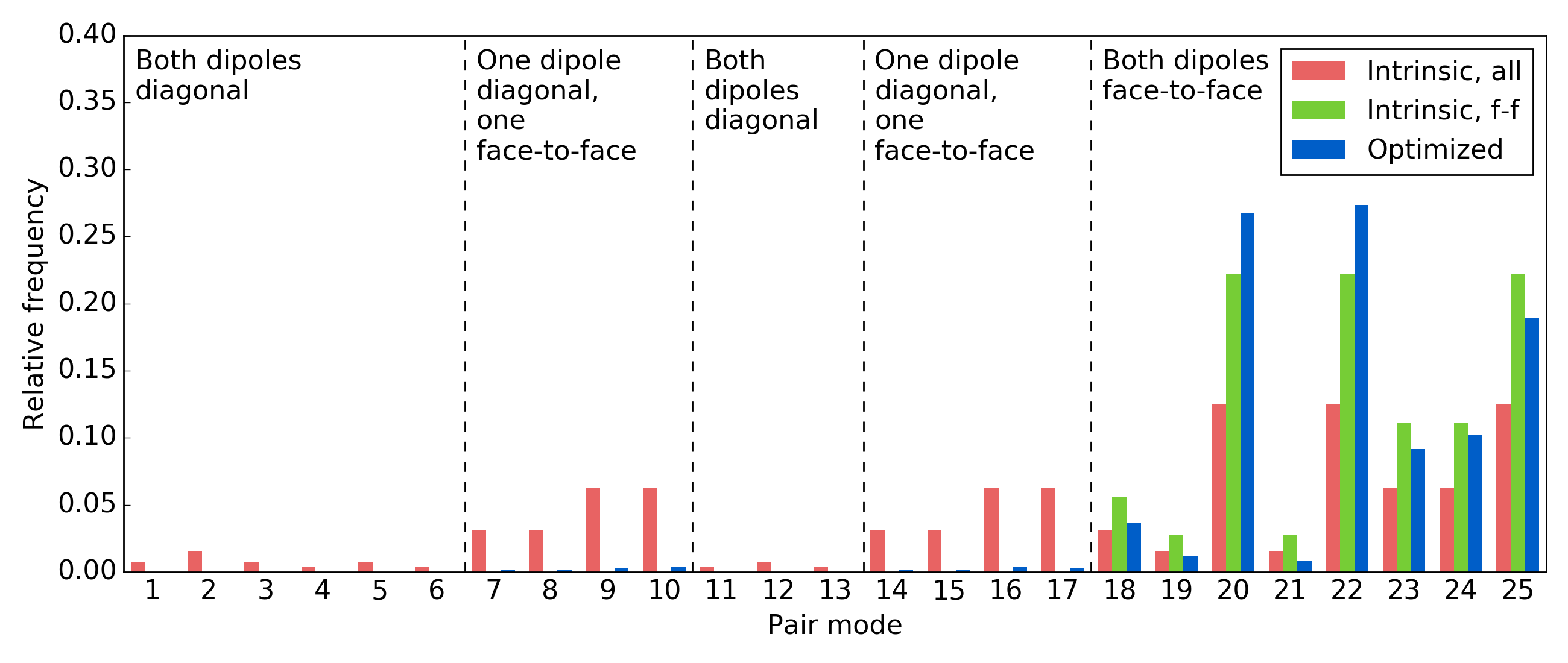} \caption{PM
    distribution of the 33 energy-optimized MAPbI$_3$ $4 \times 4 \times 4$
supercell models (blue), in comparison with the intrinsic probabilities of all
the PMs (red) and the intrinsic probabilities of only the face-to-face PMs
(green).} \label{fig_pm_distro_25} \end{figure}

In our recent study \cite{Li2017} we optimized 33 $4 \times 4 \times 4$ MAPbI$_3$ supercells (see figure~\ref{fig_coarse}(b)) with DFT and studied the effect of optimization on the PM distribution (see figure~\ref{fig_pm_distro_25}).  We observed that PMs 1--17, in which one or both dipoles are diagonal, are extremely rare in the optimized structures; much rarer than in an ``intrinsic'' random distribution (see red bars in figure~\ref{fig_pm_distro_25}). This is an obvious consequence of the higher energy of the diagonal structure. Comparing our optimized PM distribution to then only the face-to-face ``intrinsic'' PM probabilities of a random system (green data in figure~\ref{fig_pm_distro_25}), we observe a notable increase in the frequency of PMs 20 and 22. These correspond to perpendicular-aligned dipole pairs. Accordingly, the frequency of the other PMs decrease. To study the dipole order in MAPbI$_3$ on a large scale and its effect on the charge migration properties, we then generate new structures based on this optimized PM distribution.

\subsection{Large-scale structure generation} \label{struct_gen}

We generated the structures using our own algorithm, in which the dipoles are reoriented one-by-one in a random order to reach the given PM distribution (the ``target'' distribution, hereafter). In this study, the target distribution corresponds to the optimized PM distribution, shown in figure~\ref{fig_pm_distro_25}.  The process is started from an initial state, in which the dipoles are oriented randomly in the 24 face-to-face orientations.  These orientations correspond to the minimum-energy structures found in our previous study \cite{Li2016}.  New dipole orientations are then determined by evaluating the change that any of the other orientations would induce in the PM distribution.  The dipole is then rotated to the new orientation, that is the most favorable with respect to the target distribution, and the algorithm proceeds to the next random dipole. Via this process, the system evolves towards the target distribution, until a specified error threshold between the distributions has been reached.  Due to a noisy variation in the present distribution, the threshold is evaluated with respect to the mean of the 10 previous root-mean-square errors (RMSEs) between the present and the target distributions.  In practice, the structure generation algorithm consists of an iteration loop, in which the following steps are executed:

\begin{enumerate}[(i)] \setlength\itemsep{0em}
    \item Select a random dipole
    \item Rotate the dipole through all 32 orientations to see, which one
	produces the most favorable change towards the target PM distribution
    \item Reorient the dipole to the most favorable orientation
    \item Calculate the present PM distribution
    \item Calculate the RMSE between the present 
	and target distributions. Stop if the specified accuracy is reached.
    \item Return to (i)
\end{enumerate}

\noindent In this process, we also analyzed the electrostatic dipole-dipole interaction energy of the structure. The energy is

\begin{equation} \label{eq_es_energy}
  	U = - \sum_{i} \sum_{j \neq i} \bi{p}_i \cdot \bi{E}_{ij}
    = -\frac{1}{4 \pi \epsilon_0 \epsilon_r}\sum_{i} \sum_{j \neq i} \bi{p}_i \cdot \frac {3(\bi{p}_j \cdot \hat{\bi{r}}_{ij}) \hat{\bi{r}}_{ij} - \bi{p}_j}{|\bi{r}_{ij}|^3},
\end{equation}

\noindent in which $\bi{p}$ is the electric dipole moment, $\bi{E}_{ij}$ is the electric field at dipole $i$ by dipole $j$, $\bi{r}_{ij}$ is the distance vector from dipole $j$ to dipole $i$, $\epsilon_0$ is the vacuum permittivity and $\epsilon_r$ is the dielectric constant. The results of the structure generation are presented in section~\ref{res_struct_gen}.

\subsection{Hopping model for charge migration} \label{mig_model}

\begin{figure}[ht] \centering
    \includegraphics[width=1.0\textwidth]{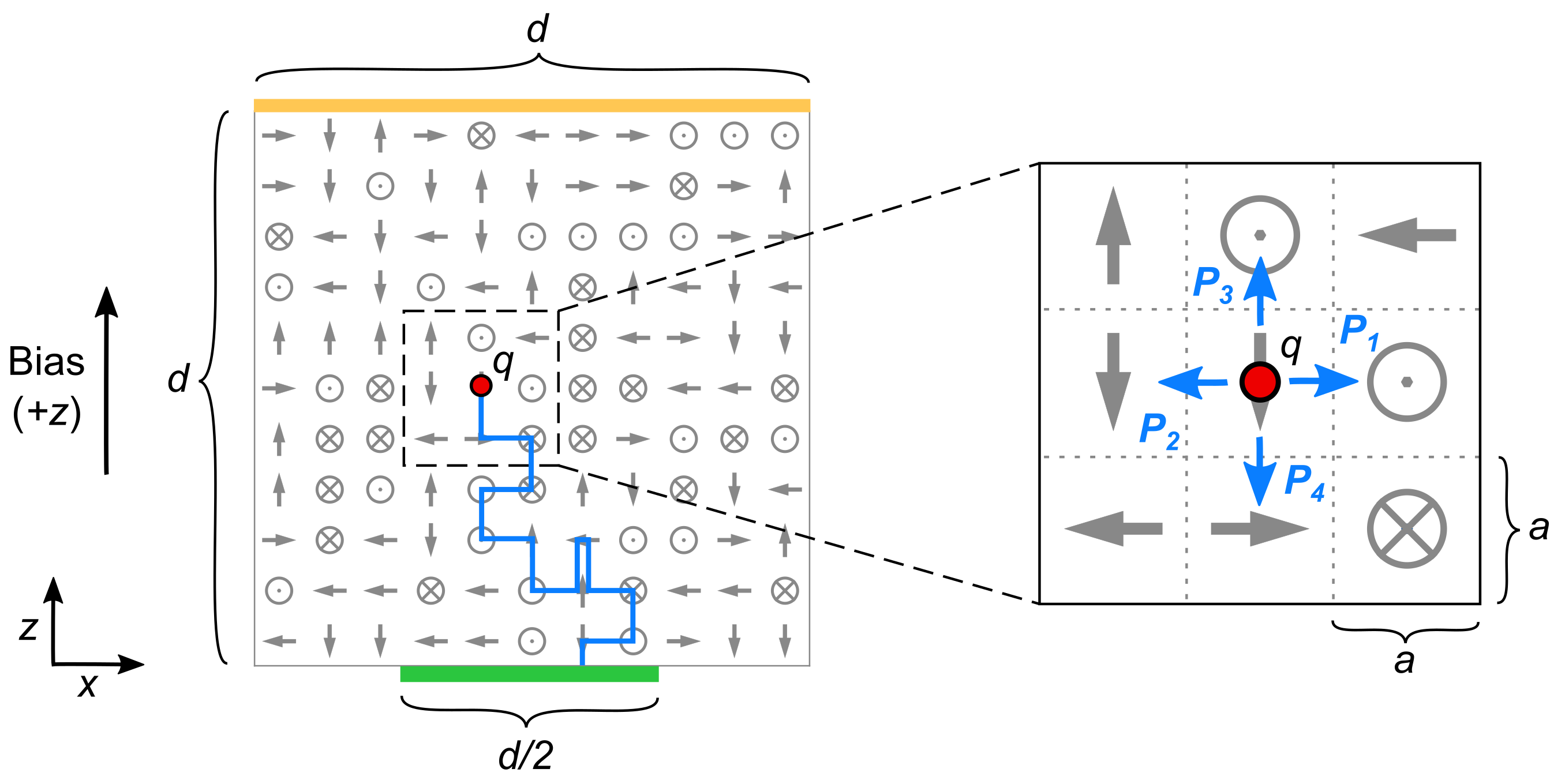} \caption{
	2-dimensional representation of the mixed quantum-classical hopping model, in
	which a charge $q$ (red) executes a classical random walk (blue)
	according to the QM hopping probabilities $P_{1-4}$, in a simulation box
	of dimension $d$ with lattice constant $a$.  The migration paths are
	initiated at the middle section of the bottom plane (green).  The
	hopping probabilities are biased in the $+z$ direction, thus driving the
	charge towards the opposite side of the simulation box (yellow).}
\label{fig_hop_diag} \end{figure}

\begin{figure}[ht] \centering
    \includegraphics[width=1.0\textwidth]{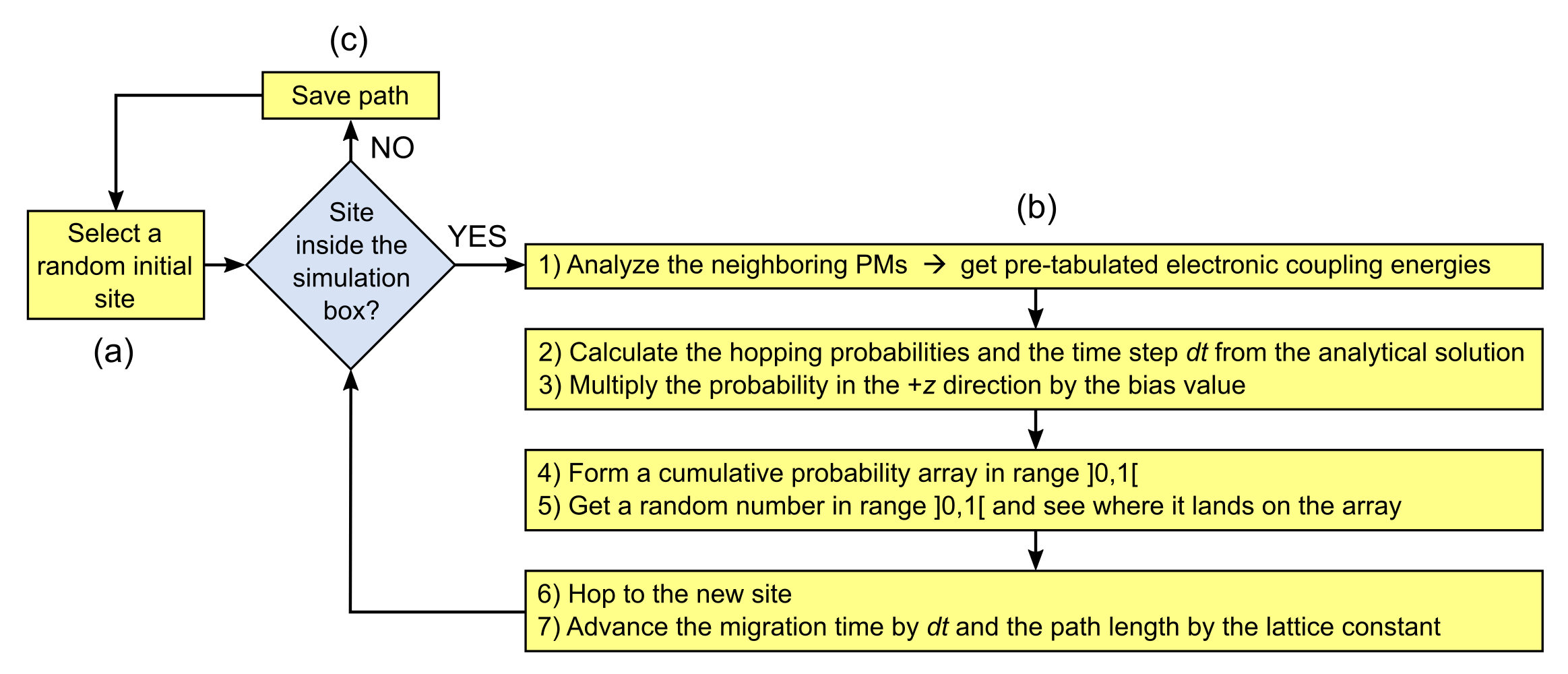}
\caption{Flow diagram of the path-hopping algorithm in the
    charge-migration analysis. Each path is (a) initialized at a random site at
    the bottom plane of the simulation box. A single hop comprises (b) a 7-stage
    process of evaluating the hopping direction and the time step. The hopping
    is repeated until (c) the path exits the simulation box, after which a new
    path is initialized.  The algorithm is executed for a pre-specified number
of paths.  } \label{fig_mig_chart} \end{figure}

We analyzed the effect of dipole order on the charge migration using a self-developed algorithm (see figures~\ref{fig_hop_diag} and \ref{fig_mig_chart}), in which a fully localized charge-carrier executes a classical random walk in a simulation box, based on PM-specific NN hopping probabilities. The QM behavior of the charge-carrier is evaluated at each hop based on an analytical solution of the hopping probabilities and the time step (see section~\ref{analytical}), without the need to compute the time evolution of the QM wave-function explicitly. The analytical solution applies the PM-specific electronic coupling energies between the NNs, which are pre-computed transfer integrals (TIs) from the DFT band structure of MAPbI$_3$ (see section~\ref{coupling}). Importantly, the analytical solution results in an extremely low computational simulation cost. This facilitates the calculation of a large number of migration paths in large systems.  For example, in this study, each sample set consists of $10^4$ paths, which were calculated in systems of $10^6$ unit cells and above.

Each hopping path was started from a random initial location at the bottom plane ($z=0$) of the system (see figure~\ref{fig_mig_chart}(a)). To increase the number of the paths that migrate through the system and not exit from the sides, the initial sites were selected from the middle half of the bottom plane (in range $[d/4,3d/4]$, in which $d$ is the dimension of the system in $x$ or $y$ direction). The hopping was then executed (see figure~\ref{fig_mig_chart}(b)) until the path exited the system, from any of the 6 sides of the system (see figure~\ref{fig_mig_chart}(c)). To simulate the force of an external electric field on the charge-carrier, we applied a constant bias multiplier to the hopping probabilities in the direction of the electric field, which was set to $+z$. The multiplier increased the number of the paths that migrated through the system, which was essential in obtaining a sufficiently large set of migration times for the statistical charge velocity.  For this model, we first derived the analytical solution for the NN hopping probabilities (see section~\ref{analytical}). The solution applies the electronic coupling energies $V_n$ between the NNs, which we calculated from the DFT band structure of MAPbI$_3$ (see section~\ref{coupling}).

\subsubsection{Analytical solution for hopping probabilities} \label{analytical}

The NN hopping probabilities and the time step have an analytical QM solution (see \ref{app_analytical} for details). The Hamiltonian consists of the NN coupling energies $V_n$ between the initial site 0 and the neighboring sites $n \in [1,6]$, such that

\begin{equation} 
    \hat{H} = \sum_{n = 1}^{6} \Big(\left|0\right> V_n \left<n\right| +
\left|n\right> V_n \left<0\right|\Big).  \end{equation}

\noindent We defined the squared sum of the coupling values as

\begin{equation}
    V = \left( \sum_{n = 1}^{6} V_n^2 \right)^{1/2}.
\end{equation}

\noindent From these, we derived the time evolution of the wave function and the probabilities for the charge to be at the initial site or at a neighboring site. The probability for the charge to be at the initial site at time $t$, starting from a fully localized state of the wave function, is

\begin{equation}
    P(0;t) = \cos^2 \left(\frac{Vt}{\hbar}\right).
\end{equation}

\noindent Importantly, the probability $P(0;t)$ is a simple function of $V$ and does not require solving the QM time evolution of the wave function explicitly in each hopping step, which is computationally expensive.  As can be seen from the cosine function, the probability for the charge to be at the initial site starts to decrease as time progresses.  Subsequently, the charge has a non-zero probability to exist at 7 different sites, i.e. at the initial site or at any of the 6 neighboring sites. Thus, we approximated that the charge will hop, on average, once its initial-site probability has decreased below 1/7. The hopping time is therefore

\begin{equation}
    t_{\rm{hop}} = \frac{\hbar}{V} \cos^{-1} \left(\frac{1}{\sqrt{7}}\right).
\end{equation}

\noindent At the time $t_{\rm{hop}}$, the probability of the charge to be at site $n$
is

\begin{equation}
    P(n;t_{\rm{hop}}) = \left(\frac{V_n}{V}\right)^2 \sin^2 \left(\frac{Vt_{\rm{hop}}}{\hbar}\right).
\end{equation}

\noindent We then applied the bias multiplier to the probability of the neighbor $n$ in the direction of the electric field and determined the site to which the charge hopped according to the thus-obtained probability distribution.  Next, we will introduce the calculation of the electronic coupling energies $V_n$ from the DFT band structure of MAPbI$_3$.

\subsubsection{Electronic coupling between neighboring unit cells}
\label{coupling}

The DFT band structure of MAPbI$_3$ (see figure~\ref{fig_tis_band}(a) and \ref{app_bands}) exhibits a cosine-type dispersion relation, which is known from the tight-binding (TB) theory \cite{Slater1954}.  Based on this, we adopted the TB model in the calculation of the electronic coupling between the neighboring unit cells. The coupling energies correspond to the TIs, from which we calculated the NN hopping probabilities in the charge-migration model via the analytical solution.

In the TB approximation, the wave function $\Psi_k$ of a particle in a periodic lattice, in which $k$ is the momentum quantum number, is written as a linear combination of the atomic wave functions $\phi_n$, in which $n$ denotes the site index. In the one-dimensional single-band TB model:

\begin{enumerate}[(i)] \setlength\itemsep{0em}
	\item ${\phi_n}$ are considered to form a complete orthonormal set, i.e.
	    $\braket{\phi_n|\phi_{n'}} = \delta_{nn'}$.
	\item All sites have the same on-site energy, i.e. $\braket{\phi_n |
	    \hat{H} | \phi_n} = \rm{constant}$. This constant can be set to 0
	    without loss of generality.
	\item Only the couplings between the NNs are non-zero, which we will denote by $-\Lambda$, i.e. $\braket{\phi_n | \hat{H} |
	    \phi_{n'}} = -\Lambda \delta_{nn'\pm 1}$.
\end{enumerate}

\noindent The one-dimensional TB Hamiltonian can thus be written as

\begin{equation}
    \hat{H} = -\Lambda \sum_{n=1}^{\infty} \Big( 
    \ket{\phi_n} \bra{\phi_{n+1}}  +  \ket{\phi_{n+1}} \bra{\phi_{n}}
    \Big).
\end{equation}

\noindent In practice, we will work on finite systems, in which the total number of sites is $N$. For this, we will apply periodic boundary conditions by coupling the first and the last ($N$th) site with $-\Lambda$, such that 

\begin{equation}
    \eqalign{
	\hat{H} &= -\Lambda \bigg[ 
    \sum_{n=1}^{N-1} \Big(
    \ket{\phi_n} \bra{\phi_{n+1}}  +  \ket{\phi_{n+1}} \bra{\phi_{n}} \\
    &+ 
    \ket{\phi_N} \bra{\phi_1}  +  \ket{\phi_1} \bra{\phi_N}
    \Big) 
    \bigg].}
\end{equation}

\noindent The solutions to this eigenvalue problem are Bloch waves $\Psi_k$, and the corresponding eigenvalues $\varepsilon_k$ are cosine functions:

\begin{eqnarray} 
    \Psi_k &= \frac{1}{\sqrt{N}}\sum_{n=1}^{N} \exp(\rmi kna) \phi_n \\
    \varepsilon_k &= -2\Lambda \cos(ka), \label {eq_cosineband}
\end{eqnarray}

\noindent in which $a$ is the distance between the NN sites. With (\ref{eq_cosineband}), we can calculate the TIs as the coupling energies $-\Lambda$ from the amplitude of the conduction band in the DFT band structure of the material.

The effective mass $m^*$ of the charge-carrier can be calculated from the curvature of the conduction band (see for example \cite{Oberhofer2017}) as

\begin{equation}
    \frac{1}{m^*} = \frac{1}{\hbar^2} \frac{\rmd^2 \epsilon_k}{\rmd k^2}.
\end{equation}

\noindent Using the cosine function of the energy bands (\ref{eq_cosineband}) we
get

\begin{equation} \label{eq_effmass}
    m^* = \frac{\hbar^2}{2a^2\Lambda},
\end{equation}

\noindent in which we have applied $ka = 0$ in the conduction-band minimum. With the effective mass, we can then calculate the mobility of the charge-carrier as

\begin{equation} \label{eq_cha_mobility}
    \mu = \frac{q \tau}{m^*} = 
    \frac{2a^2 \Lambda q \tau}{\hbar^2},
\end{equation}

\noindent in which $q$ is the charge of the charge-carrier and $\tau$ is the scattering time. The processes that affect the scattering time in MAPbI$_3$, for example the electron-phonon coupling, are outside the scope of our charge-migration model. Thus, we calculated the charge-carrier mobility based on an experimental reference value for the scattering time in the normal-temperature tetragonal phase. We obtained the mobilities for the other analyzed structures by scaling the computed velocities in our charge-migration model linearly in relation to the tetragonal-phase mobility.  The results of this analysis are presented in section~\ref{res_migration}.

\subsection{Computational Details}

We performed the band-structure DFT calculations with the all-electron numeric-atom-centered orbital code \texttt{FHI-aims} \cite{Blum2009,Havu2009,Levchenko2015,Xinguo/implem_full_author_list}.  For all calculations we used tier 2 basis sets and the PBE exchange-correlation functional \cite{Perdew1996} augmented with long-range van der Waals corrections based on the Tkatchenko-Scheffler method of Hirshfeld electron-density partitioning \cite{Tkatchenko2009}. In addition, scalar relativistic effects were included by means of the zero-order regular approximation \cite{Lenthe1993}.  A $\Gamma$-centered $4 \times 4 \times 4$ k-point mesh was used for the supercell models a and b (see \ref{app_bands}), and a $8 \times 8 \times 8$ k-point mesh for the primitive-cell model c (see figure~\ref{fig_tis_band}(a)). The results of the band-structure calculations are available from the Novel Materials Discovery (NoMaD) repository \cite{NOMADbands}.

\section{Results and discussion} \label{results}

\subsection{Large-scale dipole structures} \label{res_struct_gen}

\begin{figure}[ht] \centering
    \includegraphics[width=1.0\textwidth]{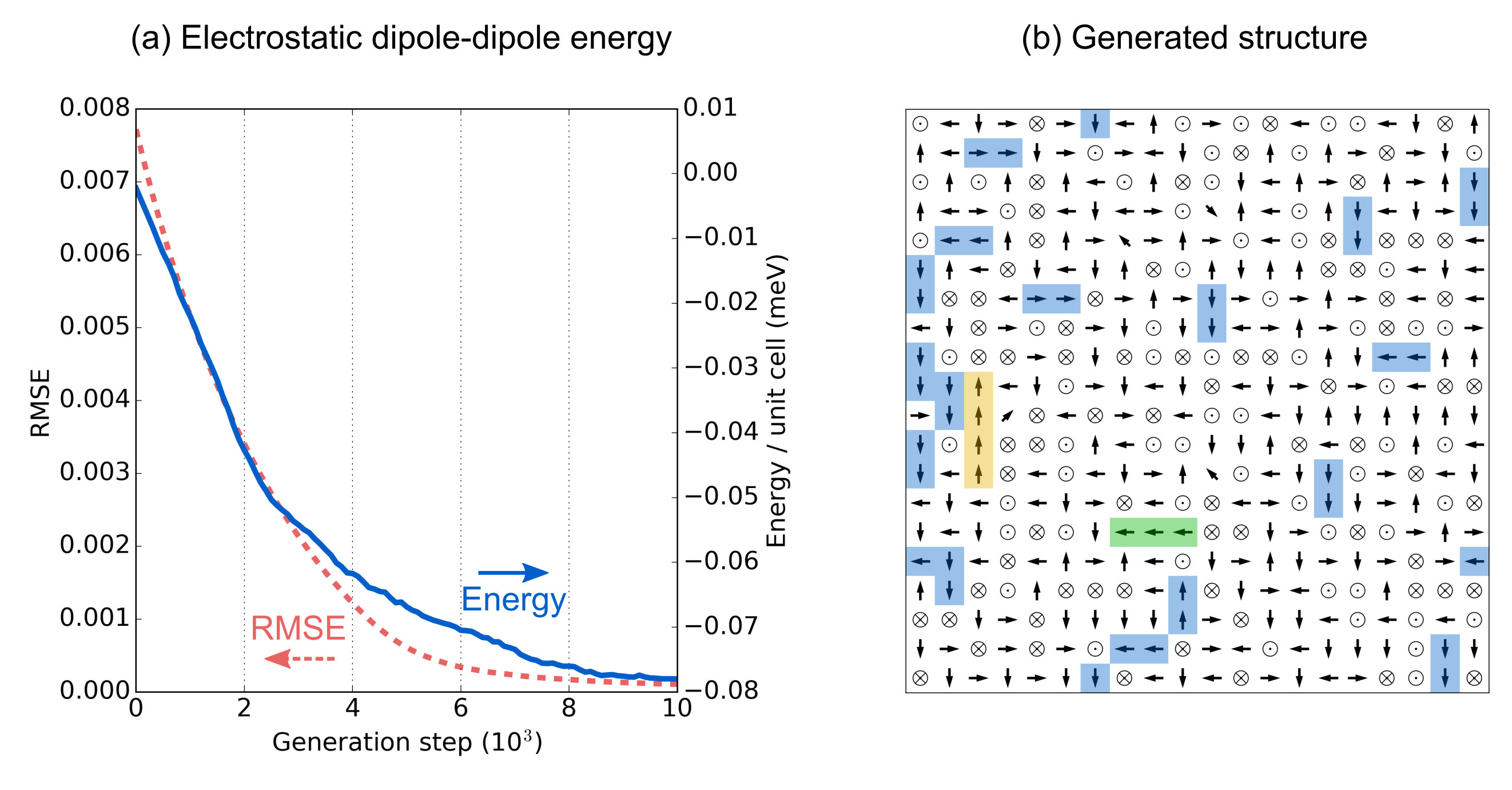}
    \caption{(a) Electrostatic dipole-dipole energy per unit cell and the
    root-mean-square error (RMSE) from the optimal pair mode distribution in the
    structure generation. The energy and RMSE are mean values of 50 generated
    $20 \times 20 \times 20$ structures, starting from initially random
    face-to-face dipole orientations, and (b) a representative $xy$-plane of the
    generated $20 \times 20 \times 20$ structures. Parallel-aligned dipole
configurations of length 2, 3 and 4 are highlighted in blue, green and yellow,
respectively.} \label{fig_structure_generation} \end{figure}

First, we studied the performance of the structure-generation algorithm by generating 50 structures of $20 \times 20 \times 20$ unit cells (see figure~\ref{fig_structure_generation}(b)).  In the process, we followed the electrostatic dipole-dipole interaction energy (\ref{eq_es_energy}) within a cutoff radius of 3 unit cells (see figure~\ref{fig_structure_generation}(a)), using the experimental dielectric constant of 25.7 \cite{Brivio2013} and the dipole moment of 2.3~D \cite{Frost2014b}.  We evaluated the convergence of the solution by following the RMSE between the present and the optimized PM distributions. The moderately small system size allowed us to update the PM distribution after each dipole reorientation. The RMSE was updated once in every 100 reorientations, and the system generation was terminated once the mean value of the 10 previous RMSEs decreased below $10^{-7}$.  

The results show that the optimized PM distribution is reached well within $10^4$ generation steps (i.e. dipole reorientations). The electrostatic energy, which is initially ca.~0 in a system of randomly oriented dipoles, decreases in the generation to ca.~$-0.08$~meV per unit cell. 

The generated structures suggest that, instead of the polarized domains (see e.g. Frost~et~al.  \cite{Frost2014a}), the dipole order in MAPbI$_3$ on a large scale may predominantly consist of non-aligned dipoles, which can be in perpendicular formations, corresponding to PMs 20 and 22, and the out-of-plane PM 25. This resembles the tetragonal-phase MAPbI$_3$ structure (see for example \cite{Lahnsteiner2016}), which however is not static at normal temperatures due to dipole reorientations \cite{Chen2015,Mosconi2014}. The dynamic structure may exhibit the tetragonal-phase dipole order in local domains on a small scale, while remaining macroscopically disordered.

\subsection{Transfer integrals} \label{res_tis}

\begin{figure}[ht] \centering
\includegraphics[width=1.0\textwidth]{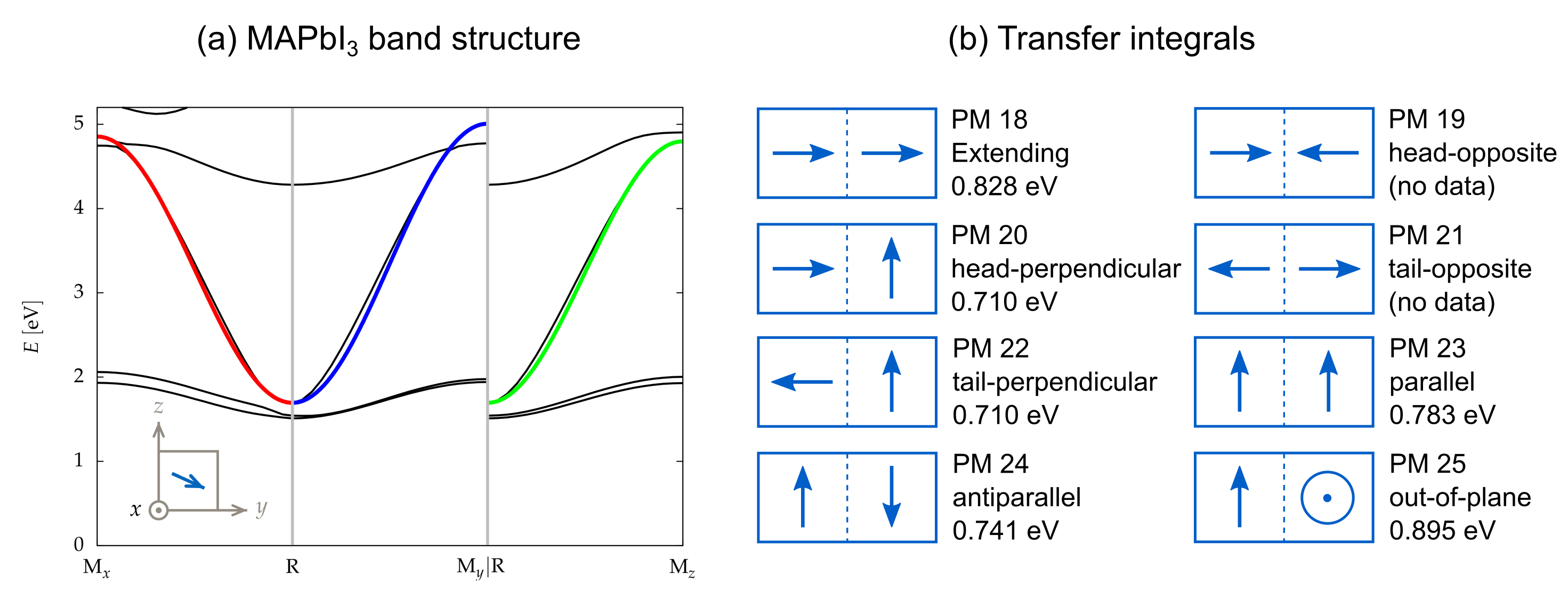}
    \caption{(a) Band structures of MAPbI$_3$ calculated with the cubic
    primitive-cell model, in which the dipoles point approximately towards $+y$
    direction (as shown in the inset). The cosine-like conduction-band
    dispersions along R-M$_x$, R-M$_y$ and R-M$_z$ are colored in red, blue and
green, respectively. (b) TIs of the face-to-face PMs 18--25, excluding the rare
    opposite-aligned PMs 19 and 21.}
\label{fig_tis_band} \end{figure}

To obtain the NN hopping probabilities in the charge-migration model, we calculated the TIs from the DFT band structure of MAPbI$_3$. We analyzed the charge migration in structures, which consist of the face-to-face PMs 18--25, thus excluding the rare diagonal dipoles. We obtained the TIs for these PMs from the band structures of one primitive-cell model and two $2 \times 2 \times 2$ supercell models. We calculated the band structure of MAPbI$_3$ in the primitive-cell model (see figure~\ref{fig_tis_band}(a)) using lattice parameters $a = b = c = 6.313$~\AA \cite{Stoumpos2013}. In the relaxed structure, the dipole is oriented approximately in the $+y$ direction. Here we use the usual notation R $\equiv (\frac{1}{2}, \frac{1}{2}, \frac{1}{2})$ for the reciprocal lattice of a simple cubic crystal system, in which the conduction-band minimum is located. In addition, we denote M$_x \equiv (0, \frac{1}{2}, \frac{1}{2})$, M$_y \equiv (\frac{1}{2}, 0, \frac{1}{2})$ and M$_z \equiv (\frac{1}{2}, \frac{1}{2}, 0)$.  The band dispersion along R-M$_{x/y/z}$ reflects the effects of the inter-site coupling along $x/y/z$. As shown in figure~\ref{fig_tis_band}(a), they all exhibit the typical cosine character and can be described with the one-dimensional TB theory. We can thus calculate the TI from the amplitude of the conduction band using (\ref{eq_cosineband}). For example, the TI of the extending PM 18 can be calculated from the primitive-cell model along R-M$_y$ as

\begin{equation}
    \Lambda_y = \frac{| E_R - E_{M_y} |}{4},
\end{equation}

\noindent in which $E_R$ and $E_{M_y}$ are the conduction band energies at points $R$ and $M_y$, respectively. Also, in these points, $ka = 0$ and $ka = \pi$ in (\ref{eq_cosineband}), respectively. To obtain the TI for the parallel-aligned PM 23, we calculated the mean value of the TIs in the primitive-cell model along R-M$_x$ and R-M$_z$. 

For the remaining PMs, we calculated the TIs from the DFT band structures of two cubic $2 \times 2 \times 2$ supercell models, which mimic the dipole orientations in the low-temperature orthorhombic phase and in the mid-temperature tetragonal phase (see \ref{app_bands}). In these supercell models, we used the lattice constant of $2a = 12.626$~\AA. In both models, the conduction band minimum is located at $\Gamma \equiv (0, 0, 0)$, which is different from the primitive-cell model due to band folding.  Instead of the common notation for the high-symmetry points in the Brillouin zone, we denote here $\rm{X} \equiv (\frac{1}{2}, 0, 0)$, $Y \equiv (0, \frac{1}{2}, 0)$ and $Z \equiv (0, 0, \frac{1}{2}$).

From the conduction band dispersion along $\Gamma$-X and $\Gamma$-Y, we calculated the TIs associated with the perpendicular PMs 20 and 22. Since both the head-perpendicular and the tail-perpendicular PMs exist in both models and cannot be singled out in the band structure, we took the mean value of the calculated TIs and assigned the same value to both PMs. Finally, the antiparallel PM 24 and the out-of-plane PM 25 correspond to the conduction band dispersion along $\Gamma$-Z in models a and b, respectively. The opposite-aligned PMs 19 and 21 are extremely rare in the energy-optimized structures (see figure~\ref{fig_pm_distro_25}) and difficult to obtain via structural relaxation in DFT.  Therefore, they were excluded from this TI analysis and assigned a coupling value of 0 eV in the charge-migration model.  The TIs for all the calculated PMs are presented in figure~\ref{fig_tis_band}(b).

\subsection{Charge migration analysis} \label{res_migration}

With our hopping model, we simulated the charge migration in various MAPbI$_3$ structures and analyzed the effect of dipole configurations on the charge mobility. We estimated the charge mobility based on the computed charge velocities, the effective mass of the charge-carrier calculated from the DFT band structure (see (\ref{eq_effmass}) and figure~\ref{fig_tis_band}(a)), and an experimental reference value for the scattering time.  Since we computed the velocities from the average charge-migration time through a finite-sized systems, they were dependent on the system size. Therefore, we first analyzed the velocities in systems of different sizes to determine the required size for a good approximation of a bulk material.

\begin{figure}[ht] \centering
    \includegraphics[width=1.0\textwidth]{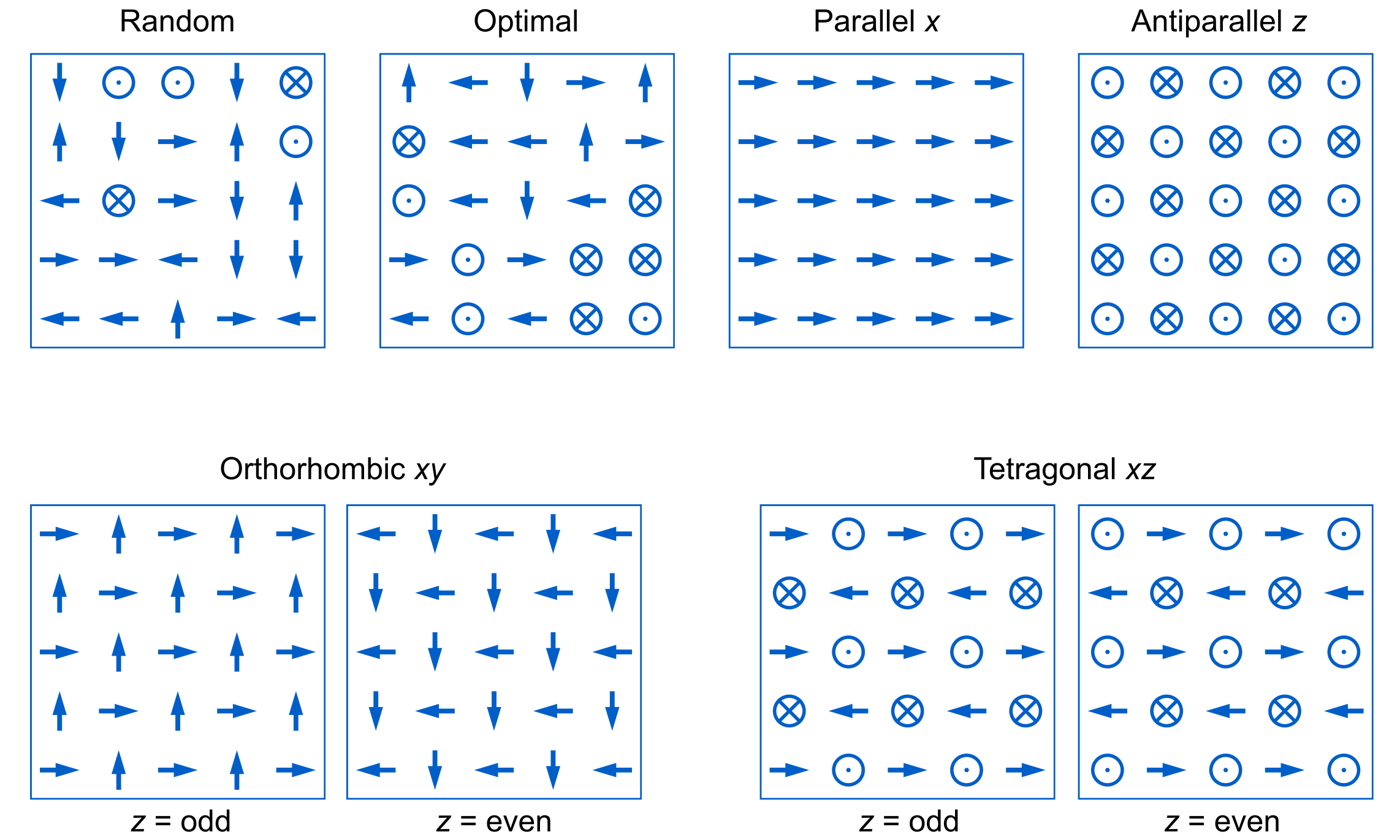}
    \caption{$5 \times 5$ unit-cell models of the basic structure types in the
    charge migration study. The models show a single $xy$ plane in the
    structure, with charge-migration in the $+z$ direction. A complete list of
the structures is presented in \ref{app_struct}.} \label{fig_structures_basic}
\end{figure}

We generated dipole configurations of 10--130 unit cells with cubic system dimensions, which are analogous to (dis)ordered MAPbI$_3$ structures and known crystal phases in the coarse-grained model. The structures (see figure~\ref{fig_structures_basic} and \ref{app_struct}) included:

\begin{enumerate}[(i)] \setlength\itemsep{0em}
	\item The low-temperature orthorhombic phase in two different
	    orientations with respect to the charge migration direction 
	    (i.e. the direction of the bias multiplier in the 
	    NN hopping probabilities)
	\item The mid-temperature tetragonal phase, in two different
	    orientations, as above
	\item Parallel and antiparallel dipoles in $z$ and $x$ directions, which correspond
	    to the parallel and perpendicular orientations with respect to the
	    charge migration direction, respectively
	\item Randomly oriented dipoles
	\item Optimized dipole configurations, generated  with the
	    structure-generation method based on the optimized PM
	    distribution
	    (see section~\ref{struct_gen}).
\end{enumerate}

\begin{figure}[ht] \centering
    \includegraphics[width=0.6\textwidth]{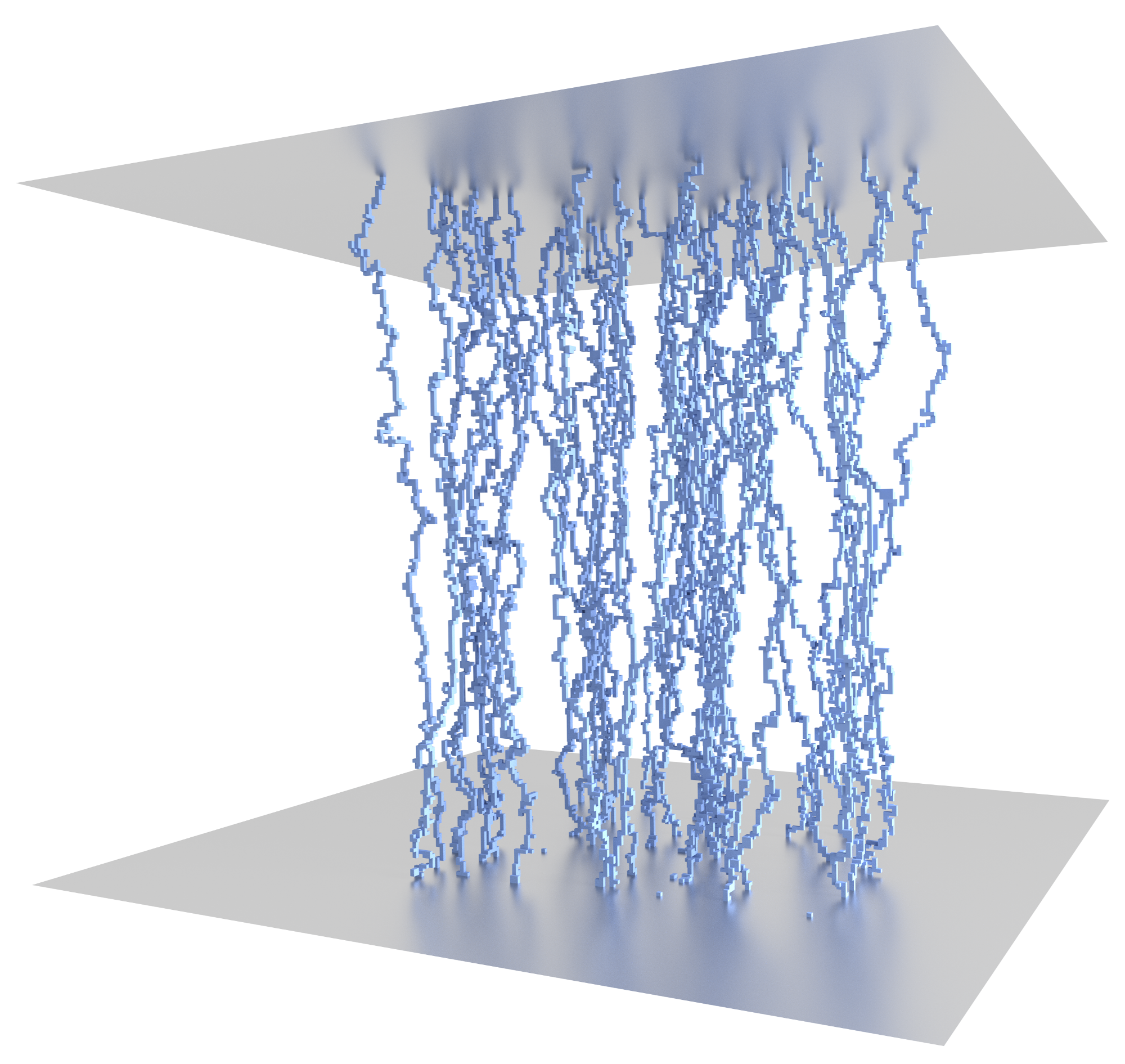} \caption{An
    example of the charge-migration paths in a system of $200 \times 200 \times
    200$ unit cells, in which 50 paths are started from the bottom plane, with
the bias multiplier set to 5 in the upwards direction.} \label{fig_paths_3d}
\end{figure}

\noindent In all the charge-migration calculations, we set the bias multiplier to 5, which resulted in ca.~80~\% of the migration paths going through the system.  Since the paths were started within the center half of the bottom plane, they rarely exited at the side-faces of the cubic system, particularly in large systems (see figure~\ref{fig_paths_3d}).  Typically, a path was terminated due to exiting through the bottom plane during the first few hops in the migration.  We computed $10^4$ paths in each of the structures, which based on test calculations on larger sample sets, provided a sufficient accuracy for the statistical velocity calculation.  

\begin{figure}[ht] \centering
    \includegraphics[width=1.0\textwidth]{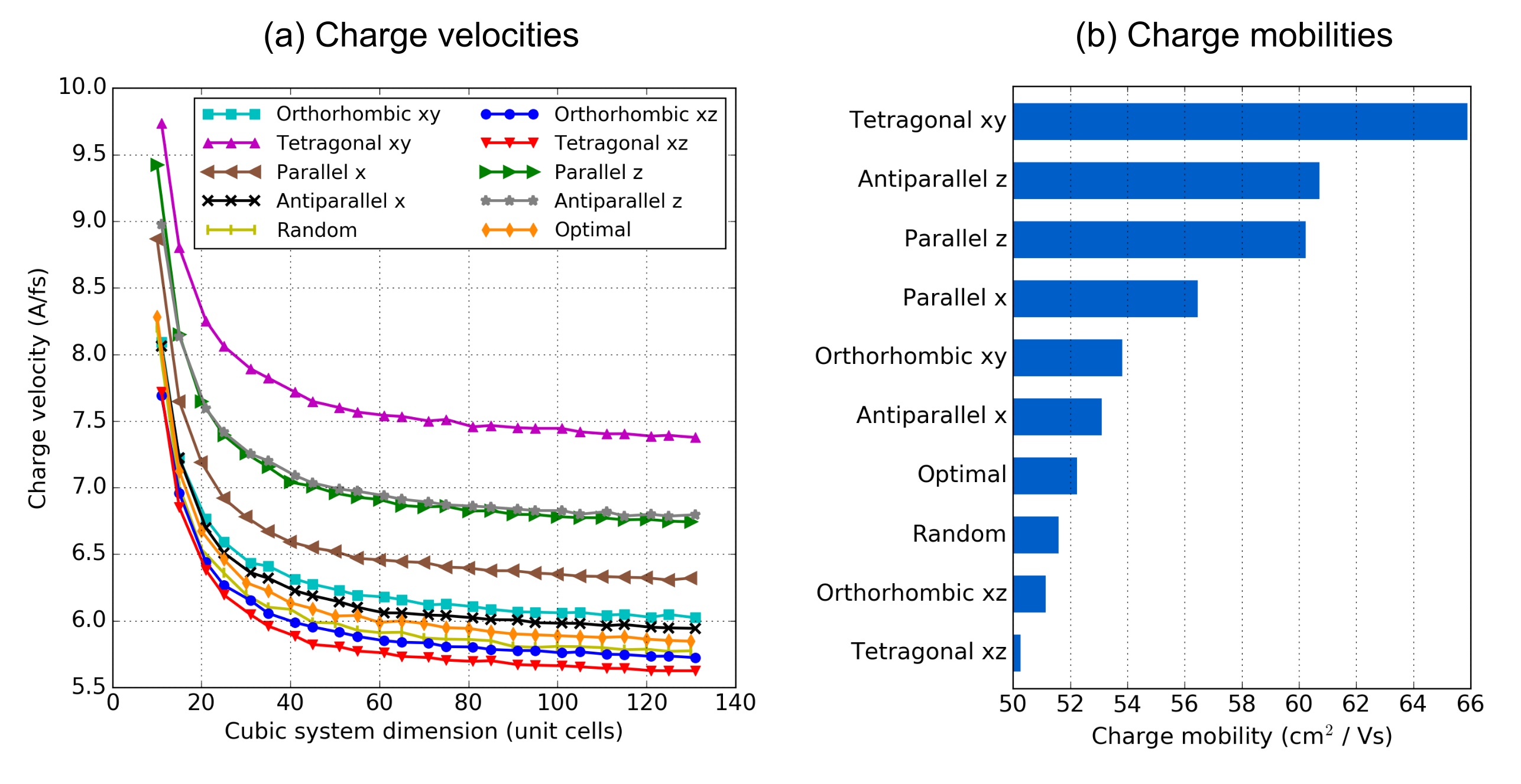} \caption{(a)
    Charge migration velocities in various dipole configurations, showing the
    dependence of the velocity on the cubic system dimension, and (b) charge
mobilities in the analyzed structures, calculated in relation to the reference
mobility of the tetragonal $xy$ structure.} \label{fig_velo_mobi} \end{figure}

The results (see figure~\ref{fig_velo_mobi}(a)) show that in all the analyzed structures, a good approximation of a bulk system was obtained with a system dimension of ca.~100 unit cells or larger. We observed the highest velocities in the mid-temperature tetragonal phase in the $xy$ direction.  With respect to the charge migration in the $+z$ direction, the tetragonal $xy$ structure consists of the out-of-plane PMs 25, which have the strongest electronic coupling, i.e.  the highest TI value (see figure~\ref{fig_tis_band}(b)).  Conversely, the tetragonal $xz$ structures had the lowest computed velocities.  This can be attributed to the low electronic coupling of the perpendicular PMs 20 and 22 in the direction of the charge migration.  In this case, the highly coupled out-of-plane PMs in the $x$ and $y$ directions effectively deviate the charge sideways, which results in a prolonged migration time through the system and thus a low velocity. The same deviation effect can be seen in the parallel and antiparallel $z$ structures, in which the higher coupling of the parallel PM 23 results in more deviation in the sideways direction and thus a lower velocity than in the antiparallel structure.  With respect to the disordered structures, the velocities were slightly higher in the generated optimized structures than in the fully random structures. This likely results from the reduced number of the non-coupling PMs 19 and 21 in the optimized structures.

\begin{figure}[ht] \centering
    \includegraphics[width=1.0\textwidth]{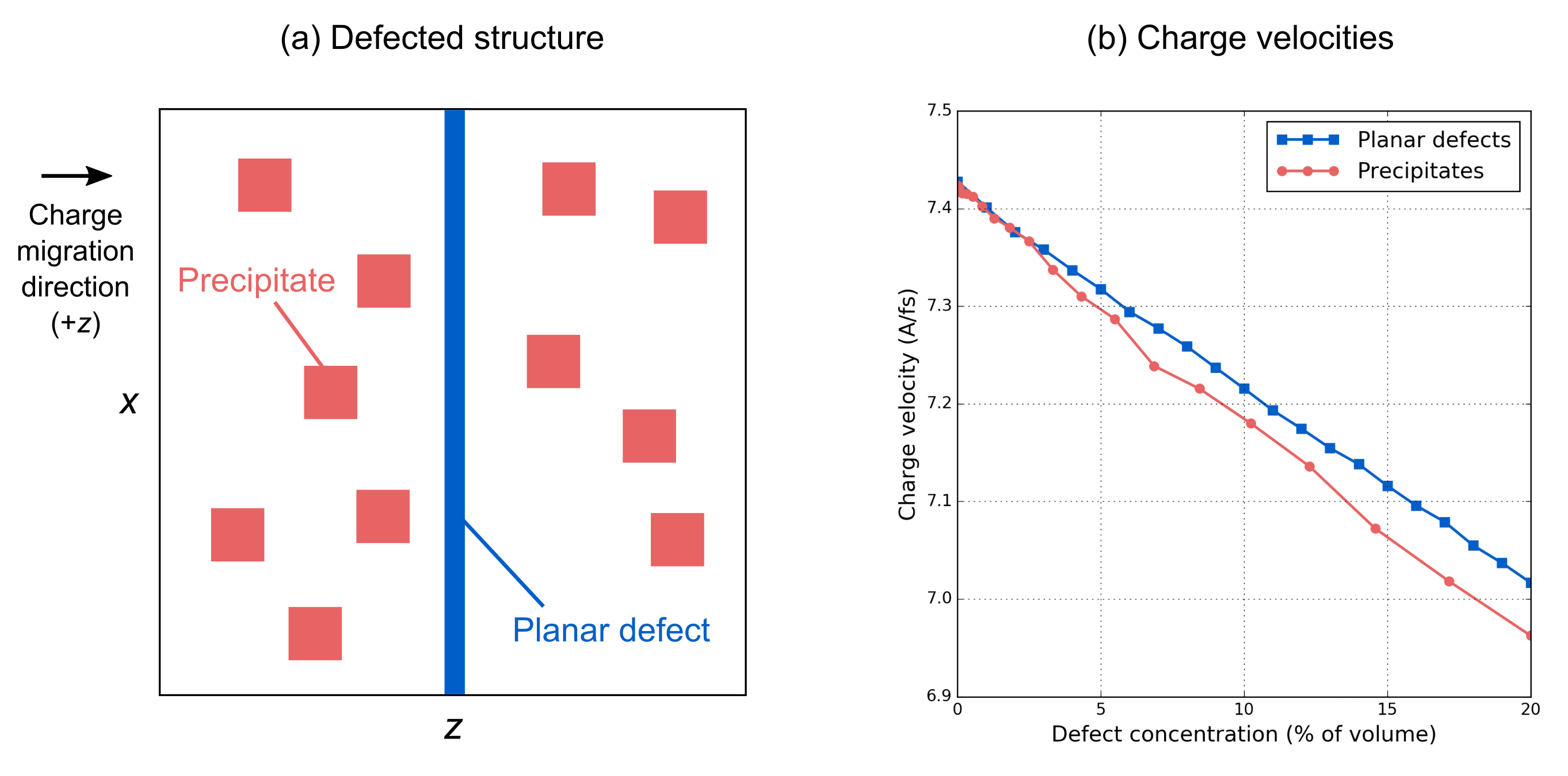}
    \caption{(a) Schematic of an $xz$-plane of a defected structure (dipoles not shown), illustrating the two defect types: a planar defect (blue) and precipitates (red). (b) Effect of the defects on the charge velocity in the tetragonal $xy$ structures of $100 \times 100 \times
100$ unit cells. The velocities are calculated as a mean value of 20 systems of
each defect concentration.} \label{fig_defect_structures} \end{figure}

We analyzed the effect of structural defects on the charge-migration velocity by introducing volumes of randomly oriented dipoles in the tetragonal $xy$ structures of $100 \times 100 \times 100$ unit cells. We studied the effect of two different types of defects: planar defects and precipitates (see figure~\ref{fig_defect_structures}(a)). In the planar defects, we introduced a system-wide plane in the middle of the system, such that the orientation of the plane ($xy$) is perpendicular to the charge migration direction ($+z$). We then varied the thickness of the plane and analyzed the charge velocity with respect to the defect volume in the system, i.e. $\widetilde{V}_{\rm{defect}} / \widetilde{V}_{\rm{system}}$.  In the precipitates, we introduced 25 randomly located non-overlapping cubic volumes of randomly oriented dipoles in the system. To reduce the noise that originates from the randomly located precipitates, we generated 20 systems of each defect concentration in 0--20 \% volume range, and computed the velocities as their mean value.

The results (see figure~\ref{fig_defect_structures}(b)) show that the velocity decreases fairly linearly with defect concentration. By extrapolating the linear velocity decrease, which starts from the tetragonal $xy$ phase velocity of ca.~7.4~\AA/fs at 0 \% defect concentration, the velocity approaches the fully random system with a velocity of ca.~5.8~\AA/fs. The linear velocity decrease is qualitatively similar in both defect types, with the precipitates having slightly lower velocities.  

Finally, we calculated the charge mobility in the analyzed structures, based on i) the computed velocities in the largest computed structures of size $130 \times 130 \times 130$ unit cells, ii) the effective mass of the charge-carrier, calculated from the tetragonal-phase DFT band-structure (\ref{eq_effmass}), and iii) an experimental reference value of the scattering time in the tetragonal-phase MAPbI$_3$. Karakus~et~al.  \cite{Karakus2015} have studied the electron-phonon scattering in MAPbI$_3$ with time-resolved terahertz spectroscopy, measuring a scattering time of ca.~4~fs in the tetragonal phase at 300~K. We associated this value with the best-performing tetragonal $xy$ structure in our charge-migration model, which consists of the out-of-plane PMs 25 with a TI of 0.895~eV. With these values and the lattice constant of $a = 6.313$~\AA, we estimated the best charge mobility in our model from (\ref{eq_cha_mobility}) as

\begin{equation} \label{eq_tetra_mobility} \mu_{\rm{tet}} = \frac{2a^2 \Lambda |q_{\rm{e}}|
\tau}{\hbar^2} = 65.865~\frac{\rm{cm}^2}{\rm{Vs}}, \end{equation}

\noindent in which $q_{\rm{e}}$ is the charge of an electron. The resulting mobility is more than twice the experimental reference mobility of ca.~27 cm$^2$V$^{-1}$s$^{-1}$, measured by Karakus~et~al. \cite{Karakus2015}. The difference may arise e.g.  from the calculation of the effective mass, which we have derived from the DFT band structure of the tetragonal-phase $2 \times 2 \times 2$ supercell model with the TB approximation. In reality, the atomic structure of MAPbI$_3$ is likely disordered on a macroscopic scale at room temperature.  Despite the simplified model in this study, the result is of the correct order of magnitude.

We obtained the mobilities for the other structures by relating the computed velocities to the tetragonal-phase mobility in (\ref{eq_tetra_mobility}).  First, we associated the tetragonal-phase mobility $\mu_{\rm{tet}}$ with the computed velocity of the corresponding structure $v_{\rm{tet}}$.  Then, we assigned mobilities for the other structures in relation

\begin{equation}
    \mu_i = \mu_{\rm{tet}} \frac{v_i}{v_{\rm{tet}}},
\end{equation}

\noindent in which $\mu_i$ is the charge mobility in structure $i$, and $v_i$ is its computed velocity in the charge-migration model. The results (see figure~\ref{fig_velo_mobi}(b)) show that the mobilities vary in a range of ca.~50--66 cm$^2$V$^{-1}$s$^{-1}$. The relative difference between the best- and worst-performing structures is ca.~30~\%. Also, the variation of 20 cm$^2$V$^{-1}$s$^{-1}$ is of the order of the experimentally measured charge mobility in the reference study by Karakus~et~al. The result suggests that the dipole order has a significant effect on the charge mobility in MAPbI$_3$. Interestingly, the highest and the lowest charge mobilities arise from the same tetragonal-phase structure, the difference being the direction of the charge migration. In MAPbI$_3$ preparation, correct orientation of the structure is therefore important to achieve optimal PV performance.

\subsection{Computation time} \label{comptime}

Overall, the computation times in this study are extremely low, measured in seconds or minutes. In comparison, calculations with the conventional DFT methods can typically take hours or days. The structure-generation algorithm (see section~\ref{struct_gen}) scales approximately as $\mathcal{O}(n^{2})$ with respect to the number of unit cells $n$. The scaling factor arises non-trivially from the way that the system size relates to reaching the target PM distribution within the set RMSE threshold. Since the PM distribution of the whole system is updated at regular intervals during the generation process, the computation time depends heavily on the system size.  As an example, a system of a cubic dimension of 40 unit cells takes approximately 1 minute to generate, whereas a system of a dimension of 50 unit cells requires ca.~3.5 minutes.

The charge-migration algorithm (see section~\ref{mig_model}) relies on the analytical solution for the NN hopping probabilities (see section~\ref{analytical}). Due to this, the process is extremely rapid, and the system size affects directly only on the path length in the migration direction.  In large systems, in which the overhead of initiating the process is negligible, the algorithm scales approximately linearly as $\mathcal{O}(n)$ with respect to the number of unit cells.  For a given system, the charge-migration analysis can be executed in a fraction of the time that it takes to generate the system.  For example, computing $10^5$ migration paths in a system of a cubic dimension of 50 unit cells (with the optimized PM distribution) takes ca.~2 seconds.  

The computation times presented here were calculated on a single core of a modern workstation processor at the time of this research.

\section{Conclusion} \label{conclusion}

In this work, we have investigated large-scale MAPbI$_3$ structures with a multi-scale model that applies the PM concept in the parametrization of the complex HP atomic structure. With the optimized PM distribution, we generated new coarse-grained MAPbI$_3$ structures to study the dipole configurations on a large scale. Different to the previously suggested polarized dipole order in MAPbI$_3$ \cite{Frost2014a}, our results indicate that the structure may exhibit local tetragonal-phase domains at room temperature, while being disordered on a macroscopic scale due to reorienting dipoles.  

We applied our coarse-grained model in the charge-migration study of MAPbI$_3$ with a semi-classical hopping model. We estimated the average charge-carrier velocities in various (dis)ordered dipole configurations, applying our analytical solution for the QM hopping probabilities and the electronic NN coupling energies from the DFT band-structure. We related the computed velocities to charge mobilities via an experimental reference value for the scattering time and the effective mass calculated from the DFT band-structure. 

We obtained charge mobilities in range of 50--66~cm$^2$V$^{-1}$s$^{-1}$, which are comparable to the experimental mobility of 27--150~cm$^2$V$^{-1}$s$^{-1}$ \cite{Karakus2015}. Importantly, both the highest and the lowest mobilities were obtained in the same tetragonal phase, the difference being the charge-migration direction (i.e. in the tetragonal-phase structures $xy$ and $xz$, respectively).  The relative variation of ca.~30~\% in the charge mobility suggests that the orientation of the dipoles has a substantial effect on the charge mobility in MAPbI$_3$.  In the defected tetragonal-phase MAPbI$_3$ structures we observed a linear velocity relation with respect to the defect concentration, with little difference between the planar defects and the precipitates.

With our model, we have aimed to advance the present understanding on the order of the organic cations in HPs and their effect on the charge mobility. In addition to MAPbI$_3$, our model is applicable to other HP compositions with an appropriate analysis of the PMs of the organic cations. Thus, the model provides a cost-efficient method to analyze the cation order and the charge mobility in numerous HPs. With the acquired knowledge, new HP compositions can be offered for experimental research, and modifications to the present materials can be suggested to boost their performance.

\ack 

We gratefully thank H.~Levard for insightful discussions in our collaboration.  We acknowledge the computing resources by the CSC-IT Center for Science (via the Project No.~ay6311) and the Aalto Science-IT project. An award of computer time was provided by the Innovative and Novel Computational Impact on Theory and Experiment (INCITE) program. Computing resources were also provided by the Argonne Leadership Computing Facility, which is a DOE Office of Science User Facility supported under Contract DE-AC02-06CH11357. This research was supported by the Academy of Finland through its Centres of Excellence Programme (2012-2014 and 2015-2017) under project numbers 251748 and 284621, as well as its Key Project Funding scheme under project number 305632.

\clearpage
\section*{Appendix}
\appendix

\section{Analytical solution for hopping probabilities} \label{app_analytical}

In the charge-migration model, the probability of a charge-carrier to hop to a neighboring site is based on the QM behavior of the particle. In the QM formalism, the state of a system is described by a complex wave function $\Psi(t)$ in Hilbert space. The time evolution of the wave-function can then be determined via the time-dependent Schr\"odinger equation (TDSE)

\begin{equation} \label{tdse}
    \hat{H} \Psi(\bi{r}, t) = \rmi \hbar \partial_t \Psi(\bi{r},t).
\end{equation}

\noindent In the charge-migration model, the immediate location of the particle after each hop is known with full certainty, which in the QM formalism corresponds to a fully localized wave function. As time progresses, the wave function expands to the neighboring sites in real space, according to its time-evolution via the TDSE (\ref{tdse}).  In the following, we will show that when considering the hopping probabilities to the NN sites, the wave function resides in a two-state Hilbert space. The corresponding two-state system has a known analytical solution, from which the hopping probabilities can be determined without explicitly calculating the time-evolution of the wave function in each hop. This is extremely valuable for the computational efficiency of the charge-migration model.

\begin{figure}[h] \centering
    \includegraphics[width=0.3\textwidth]{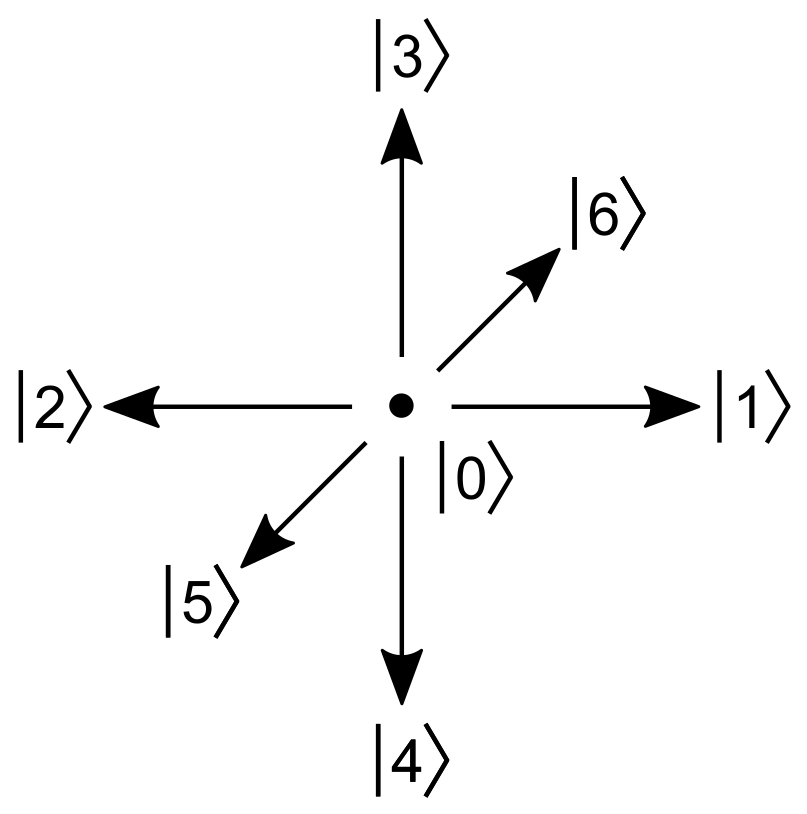}
    \caption{Localized charged states, with $\ket{0}$ corresponding to the
initial site before hopping and $\ket{1}-\ket{6}$ to the neighboring
sites.} \label{locbasis} \end{figure}

\noindent First, we construct a localized basis set

\begin{equation}
S = \{ \ket{0}; \ket{n} \},\ \{ n \in \mathbb{Z} : 1 \geq n \geq 6 \},
\end{equation}

\noindent in which states $\ket{0}$ and $\ket{n}$ correspond to the charge being located at the initial site or at any of the 6 neighboring sites, respectively (see figure~\ref{locbasis}). Here we consider that these 7 states are orthonormal and form a complete set, so that the Hamiltonian and the time-dependent wave function $\Psi(t)$ can be represented in terms of them:

\begin{eqnarray} 
    \braket{0|n} &= 0\ \forall n \in [1, 6] \\
    \braket{n|n'} &= \delta_{nn'}\ \forall n, n' \in [1, 6].  
\end{eqnarray}

\noindent The Hamiltonian consists of the NN electronic coupling energies $V_n$ between the initial site 0 and the neighboring sites $n \in [1,6]$, such that

\begin{equation}
    \hat{H} = \sum_{n=1}^{6}
    \Big( \ket{0} V_n \bra{n} + \ket{n} V_n \bra{0} \Big).
\end{equation}

\noindent Since we have a freedom of choosing an appropriate basis to represent the system, we can construct a new delocalized basis as a superposition of the localized states $\{ \ket{0}; \ket{n} \}$. The new basis set (also orthonormal) is

\begin{equation} 
    S' = \{ \ket{0}; \ket{\phi}, \ket{\varphi_k} \},
    \ \{ k \in \mathbb{Z} : 1 \geq k \geq 5 \},
\end{equation}

\noindent in which the basis states are

\begin{eqnarray}
    \ket{\phi} &= \sum_{n=1}^6 \ket{n} \braket{n|\phi}
    \equiv \sum_{n=1}^6 \alpha_n \ket{n},\ \alpha_n \in \mathbb{R}
    \label{locphi} \\
    \ket{\varphi_k} &= \sum_{n=1}^6 u_{kn} \ket{n},\ u_{kn} \in \mathbb{R}.
\end{eqnarray}

\noindent We define the coefficients $\alpha_n$ to be

\begin{equation}
    \alpha_n = \frac{V_n}{V},\ V \equiv \left( \sum_{n'=1}^{6} V_{n'}^2 \right)^{1/2},
\end{equation}

\noindent such that

\begin{equation} \label{alphasum}
    \sum_{n=1}^6 \alpha_n^2 = 
    \sum_{n=1}^6 \frac{V_n^2}{\sum_{n'=1}^6 V_{n'}^2} = 1.
\end{equation}

\noindent State $\ket{n}$ in the delocalized basis is

\begin{equation}
    \eqalign{
	\ket{n} &= \ket{\phi} \braket{\phi|n} + \sum_{k=1}^5 \ket{\varphi_k} 
    \braket{\varphi_k|n} \\
    &= 
    \alpha_n \ket{\phi} + \sum_{k=1}^5 u_{nk} \ket{\varphi_k},}
\end{equation}

\noindent and then

\begin{equation}
    \eqalign{
    \hat{H} &= \sum_{n=1}^6 
    \Bigg(  
    \ket{0} V_n \bra{\phi} \alpha_n + 
    \sum_{k=1}^5 \ket{0} V_n \bra{\varphi_k} u_{nk} \\
    &+ \alpha_n \ket{\phi} V_n \bra{0} + 
    \sum_{k=1}^5 u_{nk} \ket{\varphi_k} V_n \bra{0}
    \Bigg) \\
    &= \sum_{n=1}^6 V_n
    \Bigg[  
    \alpha_n \Big( \ket{0} \bra{\phi} + \ket{\phi} \bra{0} \Big) \\
    &+
    \sum_{k=1}^5 u_{nk} \Big( \ket{0} \bra{\varphi_k} + 
    \ket{\varphi_k} \bra{0} \Big)
    \Bigg] \\
    &= V \sum_{n=1}^6 
    \Bigg[  
    \alpha_n^2 \Big( \ket{0} \bra{\phi} + \ket{\phi} \bra{0} \Big) \\
    &+
    \sum_{k=1}^5 \alpha_n u_{nk} \Big( \ket{0} \bra{\varphi_k} + 
    \ket{\varphi_k} \bra{0} \Big)
    \Bigg].}
\end{equation}

\noindent In the first term, the sum over $\alpha_n^2$ equals 1 (\ref{alphasum}).  The second term vanishes due to the orthogonality of the states 

\begin{equation}
    \eqalign{
	\sum_{n=1}^6 \sum_{k=1}^5 \alpha_n u_{nk} &= 
    \sum_{n=1}^6 \sum_{k=1}^5 \braket{\varphi_k|n} \braket{n|\phi} \\
    &=
    \sum_{k=1}^5 \braket{\varphi_k|\phi} = 0.}
\end{equation}

\noindent Therefore, $\hat{H}$ in the delocalized basis consists of only two states

\begin{equation}
    \hat{H} = V \Big( \ket{0} \bra{\phi} + \ket{\phi} \bra{0} \Big).
\end{equation}

\noindent The eigenstates of $\hat{H}$ are

\begin{eqnarray} 
    \ket{\chi_1} &= \frac{1}{\sqrt{2}} \Big( \ket{0} + \ket{\phi} \Big) \\
    \ket{\chi_2} &= \frac{1}{\sqrt{2}} \Big( \ket{0} - \ket{\phi} \Big)
\end{eqnarray}

\noindent and their corresponding eigenvalues $E_1, E_2$ are

\begin{eqnarray}
    \eqalign{
	\hat{H}\ket{\chi_1} &=
    \frac{1}{\sqrt{2}}   V \Big( \ket{0} \bra{\phi} + \ket{\phi} \bra{0} \Big)
    \Big( \ket{0} + \ket{\phi} \Big) \\
    &=
    V \ket{\chi_1} \rightarrow E_1 = V} \\
    \eqalign{
	\hat{H}\ket{\chi_2} &=
    \frac{1}{\sqrt{2}}   V \Big( \ket{0} \bra{\phi} + \ket{\phi} \bra{0} \Big)
    \Big( \ket{0} - \ket{\phi} \Big) \\
    &=
    -V \ket{\chi_2} \rightarrow E_2 = -V.}
\end{eqnarray}

\noindent Also, the states $\ket{0}$ and $\ket{\phi}$ can be written in the energy eigenbasis as

\begin{eqnarray}
    \eqalign{
    \ket{0} &= \ket{\chi_1} \braket{\chi_1|0} + 
    \ket{\chi_2} \braket{\chi_2|0} \\
    &= \frac{1}{\sqrt{2}} \Big( \ket{\chi_1} + \ket{\chi_2} \Big)} \\
    \eqalign{
    \ket{\phi} &= \ket{\chi_1} \braket{\chi_1|\phi} + 
    \ket{\chi_2} \braket{\chi_2|\phi} \\
    &= \frac{1}{\sqrt{2}} \Big( \ket{\chi_1} - \ket{\chi_2} \Big).}
\end{eqnarray}

\noindent Next, we calculate the time evolution of the quantum state of the system, described by $\ket{\Psi(t)}$.  Initially, at time $t=0$, the charge is fully localized at the initial site, which corresponds to the state $\ket{\Psi(0)} = \ket{0}$. The time evolution of the system is determined by the time evolution operator $\hat{U}(t) = \exp(-\rmi\hat{H}t/\hbar)$, so that

\begin{equation}
    \eqalign{
	\ket{\Psi(t)} 
	&= \hat{U}(t) \ket{\Psi(0)} \\
	&= 
	\frac{1}{\sqrt{2}} \exp(-\rmi\hat{H}t/\hbar) 
	\Big( \ket{\chi_1} + \ket{\chi_2} \Big) \\
	&= \frac{1}{\sqrt{2}} \Big( \exp(-\rmi E_1t/\hbar) \ket{\chi_1}
	+ \exp(-\rmi E_2t/\hbar) \ket{\chi_2} \Big) \\
	&= \frac{1}{\sqrt{2}} \Big( \exp(-\rmi Vt/\hbar) \ket{\chi_1}
	+ \exp(\rmi Vt/\hbar) \ket{\chi_2} \Big).}
\end{equation}

\noindent In general, the probability $P$ of a QM system to be in a state $\ket{\xi}$ is calculated as the squared norm of the wave function amplitude

\begin{equation}
    P(\xi;t) = |\braket{\xi|\Psi(t)}|^2.
\end{equation}

\noindent In our case, the amplitude of the wave function at the initial site is

\begin{equation}
    \eqalign{
	\braket{0|\Psi(t)} &= \frac{1}{2} 
	\bra{0} \Bigg[ 
	\exp(-\rmi Vt/\hbar) \Big( \ket{0} + \ket{\phi} \Big) \\
	&+ \exp(\rmi Vt/\hbar) \Big( \ket{0} - \ket{\phi} \Big) 
	\Bigg] \\
	&= \frac{1}{2} \Big(\exp(\rmi Vt/\hbar) + \exp(-\rmi Vt/\hbar)\Big) = 
	\cos\Big(\frac{Vt}{\hbar}\Big).}
\end{equation}

\noindent The probability of the charge to be located at the initial site is then

\begin{equation}
    P(0;t) = |\braket{0|\Psi(t)}|^2 = \cos^2\Big(\frac{Vt}{\hbar}\Big).
    \label{initprob}
\end{equation}

\noindent Similarly, we get the probability of the charge to be located at any of the neighboring sites $n$ as

\begin{equation}
    \eqalign{
	P(n;t) &= |\braket{n|\Psi(t)}|^2 \\
	&= \Bigg| \frac{1}{2} \bra{n} \Bigg[
	    \exp(-\rmi Vt/\hbar) \Big( \ket{0} +
	 \sum_{n'=1}^6 \ket{n'} \braket{n'|\phi} 
	 \Big) \\
	&+ \exp(\rmi Vt/\hbar) \Big( \ket{0} - 
	 \sum_{n'=1}^6 \ket{n'} \braket{n'|\phi} 
	\Big) 
	    \Bigg] \Bigg|^2 \\
	&= \Big| \frac{1}{2} [
	    \exp(-\rmi Vt/\hbar) - \exp(\rmi Vt/\hbar) ]
	    \sum_{n'=1}^6 \braket{n'|\phi} 
	    \underbrace{\braket{n|n'}}_{\delta_{nn'}} \Big|^2 \\
	    &= \alpha_n^2 \sin^2 \Big(\frac{Vt}{\hbar}\Big) \\
	    &= \Big( \frac{V_n}{V}  \Big)^2 \sin^2 \Big(\frac{Vt}{\hbar}\Big),
    \label{nprob}}
\end{equation}

\noindent in which $\ket{\phi}$ has been written in the localized basis according to (\ref{locphi}). From this result we see that the probability of the charge to be located at the initial site, or any of the neighboring sites, oscillates at frequency $V/\hbar$. The initial-site probability $P(0;t)$ (\ref{initprob}) is 1 at time $t=0$, which corresponds to a fully localized charge at the initial site. The probability then starts to decay according to the squared cosine function, as the wave function propagates to the neighboring sites.  Concurrently, site $n$ probabilities $P(n;t)$ (\ref{nprob}) increase according to the squared sine function, so that the higher the coupling energy $V_n$ is, the more rapid is the increase.

Due to the probabilistic nature of the hopping, we cannot precisely determine the time period that the charge spends at each site. However, we can approximate the average time $t_{\rm{hop}}$ that it takes for the charge to hop to another site, based on the fact that the charge has 7 possible sites to be located at.  Therefore, on average, we approximate that the charge will hop once its probability to be located at the initial site has decreased below $1/7$. The hopping time can be solved from (\ref{initprob}) as

\begin{equation}
    t_{\rm{hop}} = \frac{\hbar}{V} \cos^{-1} \Big( \frac{1}{\sqrt{7}}  \Big).
\end{equation}

\noindent The probability of the charge to be located at site $n$ at the time of a hop can then be calculated from (\ref{nprob}) as $P(n;t_{\rm{hop}}) \equiv P_n$.  Based on these probabilities, we can then determine the site that the charge hops into. For this, we first form a cumulative array with values

\begin{equation}
    \{0, P_1, P_1 + P_2,\ \ldots\ , \sum_{i=1}^6 P_i \},
\end{equation}

\noindent and normalize the array into range $]0, 1[$. Then, we pick a random number $\{ r \in \mathbb{R} : 0 > r > 1\}$ and see, between which values it lands in the array. This then determines the site that the charge hops into, by correctly accounting for different probabilities at different sites. Finally, after each hop, the simulation time is advanced by the time step $t_{\rm{hop}}$.

\clearpage
\section{Coarse-grained structures} \label{app_struct}

\begin{figure}[ht] \centering
\includegraphics[width=0.8\textwidth]{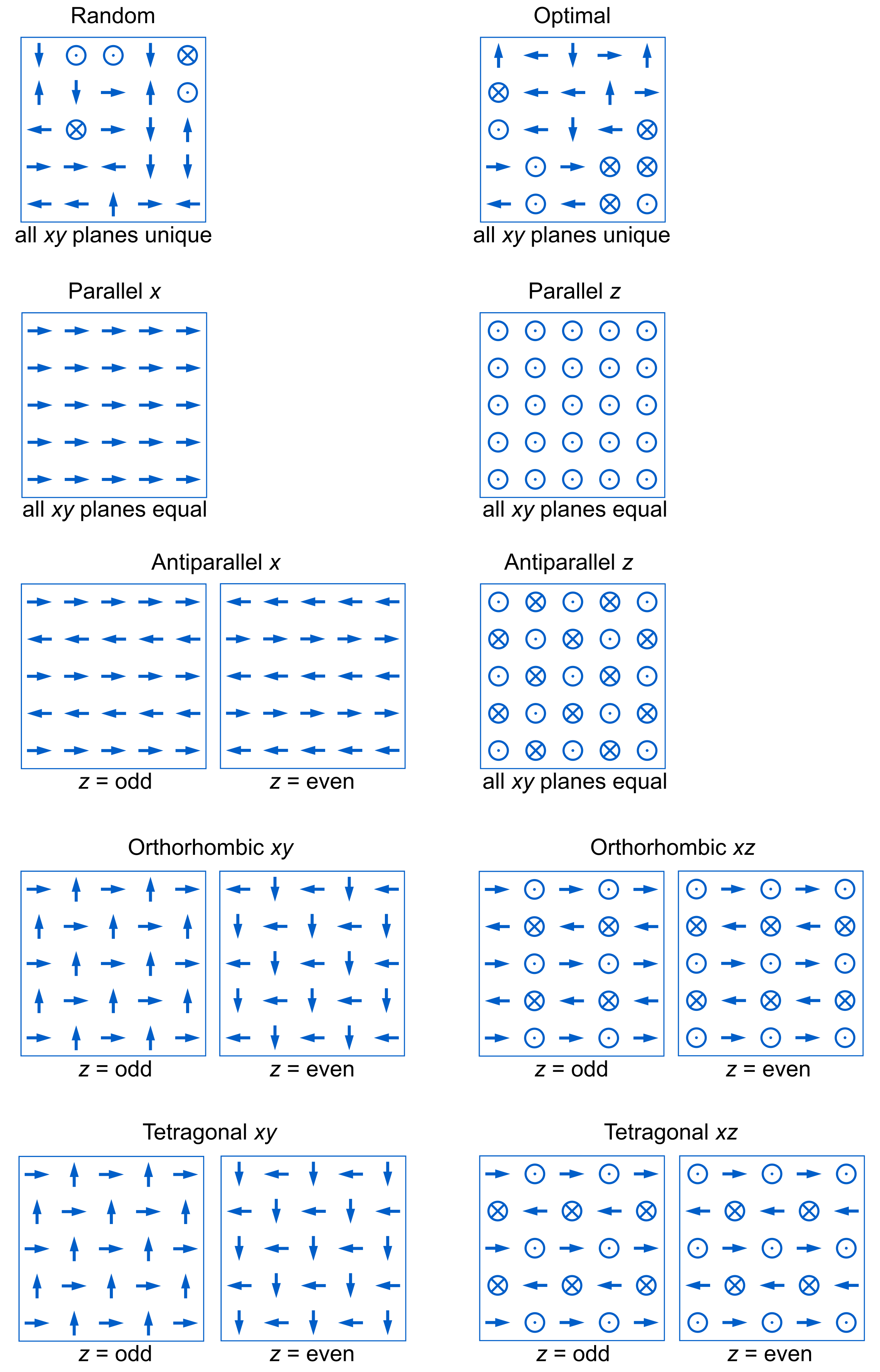}
    \caption{$5 \times 5$ unit-cell models of the structures
    in the charge-migration study. The models show a single $xy$ plane in the
    structure, with charge-migration in the $+z$ direction.}
\label{fig_structures_all} \end{figure}

\clearpage
\section{DFT band structures of MAPbI$_3$} \label{app_bands}

\begin{figure}[ht] \centering
    \includegraphics[width=0.7\textwidth]{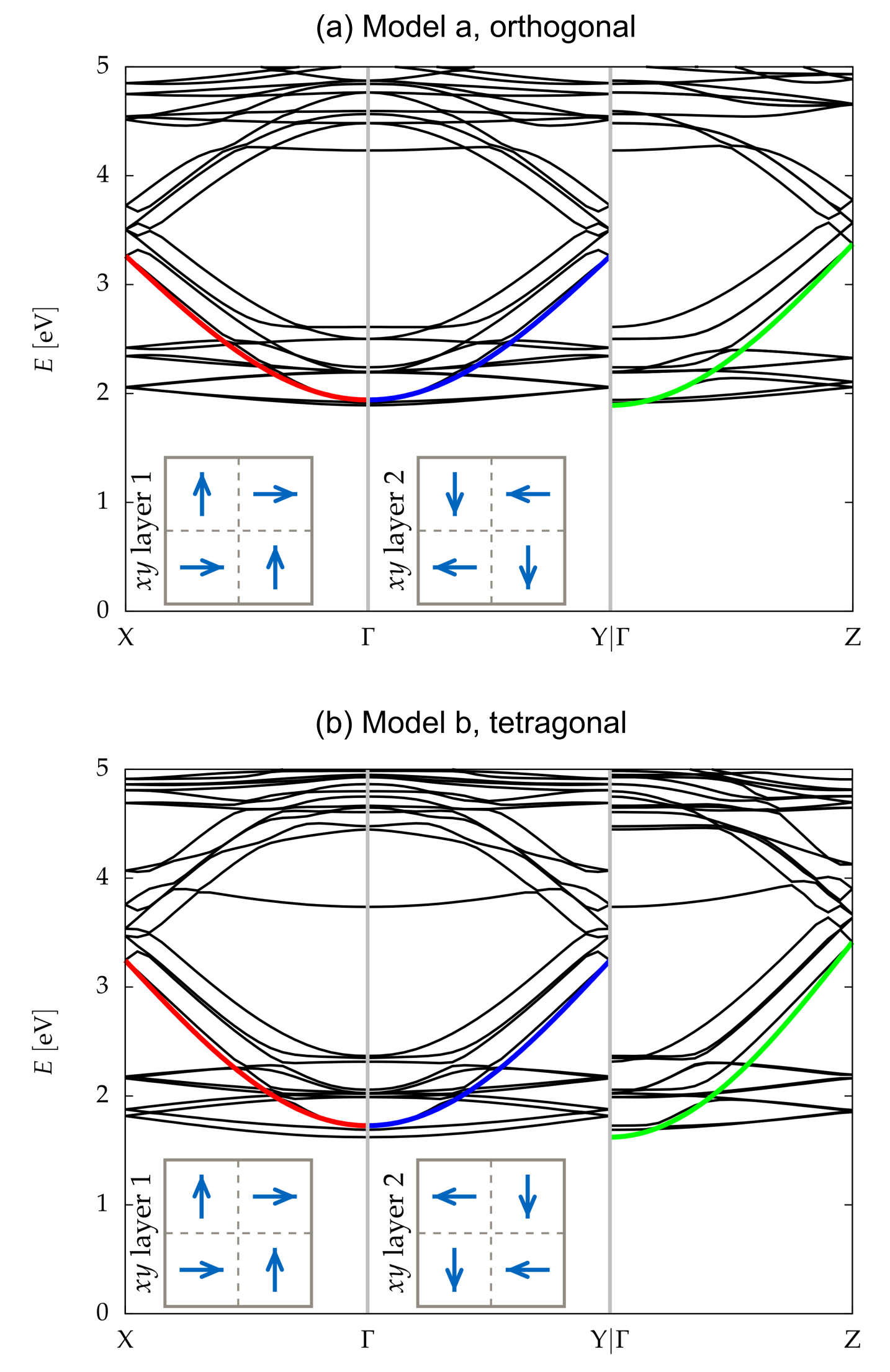}
    \caption{Band structures of MAPbI$_3$ calculated with the
    cubic $2 \times 2 \times 2$ supercell models: (a) model a, and (b) model b,
    which are analogous to the low-temperature orthogonal and mid-temperature
    tetragonal phase, respectively. For each model, the dipole orientations at
    different $xy$ planes are given as insets.  The cosine-like conduction-band
dispersions along $\Gamma$-X, $\Gamma$-Y and $\Gamma$-Z are colored in red, blue
and green, respectively.} \label{fig_bands_super} \end{figure}

\bibliographystyle{unsrt}
\clearpage
\section*{References}
\bibliography{ref}

\begin{thebibliography}{10}

\bibitem{Green2014}
M.~A. Green, A.~Ho-Baillie, and H.~J. Snaith.
\newblock The emergence of perovskite solar cells.
\newblock {\em Nature Photon.}, 8:506, 7 2014.

\bibitem{NRELchart}
National Renewable Energy Laboratory (NREL): Best research-cell efficiencies,
  2017.

\bibitem{Berhe2016}
T.~A. Berhe, W.-N. Su, C.-H. Chen, C.-J. Pan, J.-H. Cheng, H.-M. Chen, M.-C.
  Tsai, L.-Y. Chen, A.~A. Dubale, and B.-J. Hwang.
\newblock Organometal halide perovskite solar cells: degradation and stability.
\newblock {\em Energy Environ. Sci.}, 9(2):323--356, 2016.

\bibitem{Yamada2014}
Y.~Yamada, T.~Nakamura, M.~Endo, A.~Wakamiya, and Y.~Kanemitsu.
\newblock Near-band-edge optical responses of solution-processed
  organic-inorganic hybrid perovskite $\rm{CH}_3\rm{NH}_3\rm{PbI}_3$ on
  mesoporous $\rm{TiO}_2$ electrodes.
\newblock {\em Appl. Phys. Express}, 7(3):032302, 2 2014.

\bibitem{DeWolf2014}
S.~De Wolf, J.~Holovsky, S.-J. Moon, P.~L\"{o}per, B.~Niesen, M.~Ledinsky,
  F.-J. Haug, J.-H. Yum, and C.~Ballif.
\newblock Organometallic halide perovskites: Sharp optical absorption edge and
  its relation to photovoltaic performance.
\newblock {\em J. Phys. Chem. Lett.}, 5(6):1035--1039, 3 2014.

\bibitem{DongQ2015}
Q.~Dong, Y.~Fang, Y.~Shao, P.~Mulligan, J.~Qiu, L.~Cao, and J.~Huang.
\newblock Electron-hole diffusion lengths $>$175 $\mu$m in solution-grown
  $\rm{CH}_3\rm{NH}_3\rm{PbI}_3$ single crystals.
\newblock {\em Science}, 347(6225):967--970, 1 2015.

\bibitem{Brenner2015}
T.~M. Brenner, D.~A. Egger, A.~M. Rappe, L.~Kronik, G.~Hodes, and D.~Cahen.
\newblock Are mobilities in hybrid organic–inorganic halide perovskites
  actually “high”?
\newblock {\em The Journal of Physical Chemistry Letters}, 6(23):4754--4757,
  2015.
\newblock PMID: 26631359.

\bibitem{Karakus2015}
M.~Karakus, S.~A. Jensen, F.~D'Angelo, D.~Turchinovich, M.~Bonn, and
  E.~C{\'{a}}novas.
\newblock Phonon-electron scattering limits free charge mobility in
  methylammonium lead iodide perovskites.
\newblock {\em J. Phys. Chem. Lett.}, 6(24):4991--4996, 12 2015.

\bibitem{Mosconi2014}
E.~Mosconi, C.~Quarti, T.~Ivanovska, G.~Ruani, and F.~de~Angelis.
\newblock Structural and electronic properties of organo-halide lead
  perovskites: A combined ir-spectroscopy and \textit{ab initio} molecular
  dynamics investigation.
\newblock {\em Phys. Chem. Chem. Phys.}, 16:16137, 8 2014.

\bibitem{Frost2014b}
J.~M. Frost, K.~T. Butler, F.~Brivio, C.~H. Hendon, M.~van Schilfgaarde, and
  A.~Walsh.
\newblock Atomistic origins of high-performance in hybrid halide perovskite
  solar cells.
\newblock {\em Nano Lett.}, 14:2584, 5 2014.

\bibitem{Chen2015}
T.~Chen, B.~J. Foley, B.~Ipek, M.~Tyagi, J.~R.~D. Copley, C.~M. Brown, J.~J.
  Choi, and S.-H. Lee.
\newblock Rotational dynamics of organic cations in the
  $\rm{CH}_3\rm{NH}_3\rm{PbI}_3$ perovskite.
\newblock {\em Phys. Chem. Chem. Phys.}, 17(46):31278--31286, 2015.

\bibitem{Leguy2015}
A.~M.~A. Leguy, J.~M. Frost, A.~P. McMahon, V.~{Garcia~Sakai}, W.~Kochelmann,
  C.~H. Law, X.~Li, F.~Foglia, A.~Walsh, B.~C. O'Regan, J.~Nelson, J.~T.
  Cabral, and P.~R.~F. Barnes.
\newblock The dynamics of methylammonium ions in hybrid organic-inorganic
  perovskite solar cells.
\newblock {\em Nature Comm.}, 6:7124, 5 2015.

\bibitem{Lahnsteiner2016}
J.~Lahnsteiner, G.~Kresse, A.~Kumar, D.~D. Sarma, C.~Franchini, and M.~Bokdam.
\newblock Room-temperature dynamic correlation between methylammonium molecules
  in lead-iodine based perovskites: An ab initio molecular dynamics
  perspective.
\newblock {\em Phys. Rev. B}, 94(21):214114, 12 2016.

\bibitem{Li2016}
J.~Li and P.~Rinke.
\newblock Atomic structure of metal-halide perovskites from first principles:
  The chicken-and-egg paradox of the organic-inorganic interaction.
\newblock {\em Phys. Rev. B}, 94(4):045201, 7 2016.

\bibitem{Li2017}
J.~Li, J.~J\"arvi, and P.~Rinke.
\newblock Multi-scale model for disordered hybrid perovskites: the concept of
  organic cation pair modes.
\newblock {\em Phys. Rev. B, submitted, arXiv:1703.10464}, 5 2018.

\bibitem{Li2018a}
J.~Li, M.~Bouchard, P.~Reiss, D.~Aldakov, S.~Pouget, R.~Demadrille,
  C.~Aumaitre, B.~Frick, D.~Djurado, M.~Rossi, and P.~Rinke.
\newblock Activation energy of organic cation rotation in
  {CH}$_3${NH}$_3${PbI}$_3$ and {CD}$_3${NH}$_3${PbI}$_3$: Quasi-elastic
  neutron scattering measurements and first-principles analysis including
  nuclear quantum effects.
\newblock {\em J.~Phys.~Chem.~Lett., submitted}, 2018.

\bibitem{Li2018b}
J.~Li and P.~Rinke.
\newblock The motion of a methylammonium cation in the lead-triiodide cage: A
  first-principles study of hybrid perovskites using a minimal-interaction
  model. in preparation.
\newblock 2018.

\bibitem{Frost2014a}
J.~M. Frost, K.~T. Butler, and A.~Walsh.
\newblock Molecular ferroelectric contributions to anomalous hysteresis in
  hybrid perovskite solar cells.
\newblock {\em APL Mater.}, 2:081506, 8 2014.

\bibitem{Kang2017}
B.~Kang and K.~Biswas.
\newblock Preferential {CH}$_3${NH}$_3^+$ alignment and octahedral tilting
  affect charge localization in cubic phase {CH}$_3${NH}$_3${PbI}$_3$.
\newblock {\em J. Phys.: Condens. Matter}, 121(15):8319--8326, 4 2017.

\bibitem{Ma2017}
J.~Ma and L.-W. Wang.
\newblock The nature of electron mobility in hybrid perovskite
  {CH}$_3${NH}$_3${PbI}$_3$.
\newblock {\em Nano Lett.}, 17(6):3646--3654, 5 2017.

\bibitem{Hohenberg1964}
P.~Hohenberg and W.~Kohn.
\newblock Inhomogeneous electron gas.
\newblock {\em Phys. Rev.}, 136(3B):B864--B871, 11 1964.

\bibitem{Blum2009}
V.~Blum, R.~Gehrke, F.~Hanke, P.~Havu, V.~Havu, X.~Ren, K.~Reuter, and
  M.~Scheffler.
\newblock Ab initio molecular simulations with numeric atom-centered orbitals.
\newblock {\em Comput. Phys. Comm.}, 180(11):2175--2196, 11 2009.

\bibitem{Havu2009}
V.~Havu, V.~Blum, P.~Havu, and M.~Scheffler.
\newblock Efficient integration for all-electron electronic structure
  calculation using numeric basis functions.
\newblock {\em J. Comput. Phys}, 228(22):8367--8379, 12 2009.

\bibitem{Levchenko2015}
S.~V. Levchenko, X.~Ren, J.~Wieferink, R.~Johanni, P.~Rinke, V.~Blum, and
  M.~Scheffler.
\newblock Hybrid functionals for large periodic systems in an all-electron,
  numeric atom-centered basis framework.
\newblock {\em Comput. Phys. Comm.}, 192:60--69, 7 2015.

\bibitem{Xinguo/implem_full_author_list}
Xinguo Ren, Patrick Rinke, Volker Blum, J\"urgen Wieferink, Alexandre
  Tkatchenko, Sanfilippo Andrea, Karsten Reuter, Volker Blum, and Matthias
  Scheffler.
\newblock Resolution-of-identity approach to hartree-fock, hybrid density
  functionals, rpa, mp2, and gw with numeric atom-centered orbital basis
  functions.
\newblock {\em New J. Phys.}, 14:053020, 2012.

\bibitem{Perdew1996}
J.~P. Perdew, K.~Burke, and M.~Ernzerhof.
\newblock Generalized gradient approximation made simple.
\newblock {\em Phys. Rev. Lett.}, 77(18):3865--3868, 10 1996.

\bibitem{Tkatchenko2009}
A.~Tkatchenko and M.~Scheffler.
\newblock Accurate molecular van der waals interactions from ground-state
  electron density and free-atom reference data.
\newblock {\em Phys. Rev. Lett.}, 102(7):073005, 2 2009.

\bibitem{Slater1954}
J.~C. Slater and G.~F. Koster.
\newblock Simplified $\rm{LCAO}$ method for the periodic potential problem.
\newblock {\em Phys. Rev.}, 94:1498--1524, 6 1954.

\bibitem{Oberhofer2017}
H.~Oberhofer, K.~Reuter, and J.~Blumberger.
\newblock Charge transport in molecular materials: An assessment of
  computational methods.
\newblock {\em Chem. Rev.}, 117(15):10319--10357, 6 2017.

\bibitem{Lenthe1993}
E.~van Lenthe, E.~J. Baerends, and J.~G. Snijders.
\newblock Relativistic regular two-component hamiltonians.
\newblock {\em J. Chem. Phys.}, 99(6):4597--4610, 9 1993.

\bibitem{NOMADbands}
See http://dx.doi.org/10.17172/NOMAD/2018.06.08-1.

\bibitem{Brivio2013}
F.~Brivio, A.~B. Walker, and A.~Walsh.
\newblock Structural and electronic properties of hybrid perovskites for
  high-efficiency thin-film photovoltaics from first-principles.
\newblock {\em APL Mater.}, 1(4):042111, 10 2013.

\bibitem{Stoumpos2013}
C.~C. Stoumpos, C.~D. Malliakas, and M.~G. Kanatzidis.
\newblock Semiconducting tin and lead iodide perovskites with organic cations:
  Phase transitions, high mobilities, and near-infrared photoluminescent
  properties.
\newblock {\em Inorg. Chem.}, 52:9019, 8 2013.

\end{thebibliography}

\end{document}